 \documentclass[12pt,preprint]{aastex}

\usepackage{epsfig}

\newcommand{\vex}{v_{\rm ex}}

\def\green{f_{_{\rm G}}}
\def\ngreen{n_{_{\rm G}}}
\def\ugreen{U_{_{\rm G}}}

\newcommand{\vff}{v_{\rm ff}}

\newcommand{\tautrap}{\tau_{\rm trap}}
\newcommand{\ztrap}{z_{\rm trap}}
\newcommand{\zst}{z_{\rm sp}}

\newcommand{\gapprox}{\lower.4ex\hbox{$\;\buildrel >\over{\scriptstyle\sim}\;$}}
\newcommand{\lapprox}{\lower.4ex\hbox{$\;\buildrel <\over{\scriptstyle\sim}\;$}}
\newcommand{\begeq}{\begin{equation}}
\newcommand{\fineq}{\end{equation}}
\newcommand{\msun}{M_\odot} 
\newcommand{\Msun}{M_\odot} 
\newcommand{\xlum}{L_{\rm X}}
\newcommand{\chimin}{\chi_{\rm min}}
\newcommand{\chimax}{\chi_{\rm max}}
\newcommand{\epsmin}{\epsilon_{\rm min}}
\newcommand{\epsmax}{\epsilon_{\rm max}}
\newcommand{\Tmound}{T_{\rm th}}
\newcommand{\rhomound}{\rho_{\rm th}}
\newcommand{\vmound}{v_{\rm th}}
\newcommand{\zmound}{z_{\rm th}}
\newcommand{\taumound}{\tau_{\rm th}}
\newcommand{\epsilonabs}{\epsilon_{\rm abs}}
\newcommand{\chiabs}{\chi_{\rm abs}}
\newcommand{\zmax}{z_{\rm max}}
\newcommand{\taumax}{\tau_{\rm max}}
\newcommand{\deltapar}{\delta}

\def\ellprime0{\ell'_0}

\def\sig{\sigma_{_{\rm T}}}
\def\sigpar{\sigma_\|}
\def\sigperp{\sigma_\perp}
\def\sigbar{\overline\sigma}

\def\tauperp{\tau_\perp}

\def\colrad{r_0}
\def\starad{R_*}
\def\starmass{M_*}

\def\greenphoton{\dot N_\epsilon^{\rm G}}

\def\greencolumn{\Phi_\epsilon^{\rm G}}

\def\columntotal{\Phi_\epsilon^{\rm tot}}

\slugcomment{accepted for publication in \apj}

\shorttitle{X-Ray Pulsar Spectral Formation}
\shortauthors{Becker \& Wolff}

\begin{document}

\title{THERMAL AND BULK COMPTONIZATION \break
IN ACCRETION-POWERED X-RAY PULSARS}

\author{Peter A. Becker\altaffilmark{1}$^,$\altaffilmark{2}}

\affil{Center for Earth Observing and Space Research, \break
George Mason University \break
Fairfax, VA 22030-4444, USA}

\and

\author{Michael T. Wolff\altaffilmark{3}}
\affil{E. O. Hulburt Center for Space Research \break
Naval Research Laboratory, \break
Washington, DC 20375, USA}

\vfil

\altaffiltext{1}{pbecker@gmu.edu}
\altaffiltext{2}{also Department of Physics and Astronomy,
George Mason University, Fairfax, VA 22030-4444, USA}
\altaffiltext{3}{michael.wolff@nrl.navy.mil}

\begin{abstract}
We develop a new theoretical model for the spectral formation process in
accretion-powered X-ray pulsars based on a detailed treatment of the
bulk and thermal Comptonization occurring in the accreting, shocked gas.
A rigorous eigenfunction expansion method is employed to obtain the
analytical solution for the Green's function describing the scattering
of radiation injected into the column from a monochromatic source
located at an arbitrary height above the stellar surface. The emergent
spectrum is calculated by convolving the Green's function with source
terms corresponding to bremsstrahlung, cyclotron, and blackbody
emission. The energization of the photons in the shock, combined with
cyclotron absorption, naturally produces an X-ray spectrum with a
relatively flat continuum shape and a high-energy quasi-exponential
cutoff. We demonstrate that the new theory successfully reproduces the
phase-averaged spectra of the bright pulsars Her~X-1, LMC~X-4, and
Cen~X-3. In these luminous sources, it is shown that the emergent
spectra are dominated by Comptonized bremsstrahlung emission.

\end{abstract}



\section{INTRODUCTION}

More than 100 pulsating X-ray sources have been discovered in the Galaxy
and the Magellanic Clouds since the seminal detection of Her X-1 and Cen
X-3 more than three decades ago (Giacconi et al. 1971; Tananbaum et al.
1972). These sources display luminosities in the range $\xlum \sim
10^{34-38}{\ \rm ergs \ s^{-1}}$ and pulsation periods $0.1 {\ \rm s}
\lapprox P \lapprox 10^3 {\ \rm s}$, and comprise a variety of objects
powered by rotation or accretion, as well as several anomalous X-ray
pulsars whose fundamental energy generation mechanism is currently
unclear. The brightest pulsars are accretion-powered sources located in
binary systems, with emission fueled by mass transfer from the
``normal'' companion onto one or both of the neutron star's magnetic
poles. The accretion flow is channeled by the strong magnetic field into
a columnar geometry, and the resulting emission is powered by the
conversion of gravitational potential energy into kinetic energy, which
escapes from the column in the form of X-rays as the gas decelerates
through a radiative shock and settles onto the stellar surface. The
spectra of the accretion-powered sources are often well fitted using a
combination of a power-law spectrum in the $5 - 20\,$keV energy range
plus a blackbody component with a temperature in the range $T \sim
10^{6-7}\,$K (e.g., Coburn et al. 2002; di Salvo et al. 1998; White et
al. 1983) and a quasi-exponential cutoff at energy $E \sim 20-30\,$keV.
There are also indications of cyclotron features and iron emission lines
in a number of sources (e.g., Pottschmidt et al. 2005). The observations
suggest typical magnetic field strengths $\sim 10^{11-12}\,$G.

Previous attempts to calculate the spectra of accretion-powered X-ray
pulsars based on static or dynamic theoretical models have generally
yielded results that do not agree very well with the observed profiles
(e.g., M\'esz\'aros \& Nagel 1985a,b; Nagel 1981; Yahel 1980; Klein et
al. 1996). Due to the lack of a fundamental physical model, most X-ray
pulsar spectral data have traditionally been fitted using multicomponent
functions of energy that include absorbed power-laws, cyclotron
features, iron emission lines, blackbody components, and high-energy
exponential cutoffs. The resulting model parameters are often difficult
to relate to the physical properties of the source. However, the
situation improved recently with the development by Becker \& Wolff
(2005a,b) of a new model for the spectral formation process based on the
``bulk'' or ``dynamical'' Comptonization (i.e., first-order Fermi
energization) of photons due to collisions with the rapidly compressing
gas in the accretion column . In the ``pure'' bulk Comptonization model
of Becker \& Wolff, the transfer of the proton kinetic energy to the
photons, essential for producing the emergent spectrum, is rigorously
modeled using a transport equation that includes a first-order Fermi
term that accounts for the energy gain experienced by photons as they
scatter back and forth across the accretion shock. The seed photons for
the upscattering process are provided by the blackbody radiation
injected at the surface of the dense ``thermal mound,'' located at the
bottom of the accretion column (Davidson 1973). By ignoring the effect
of thermal Comptonization in the column, the model corresponds
physically to the accretion of a gas with thermal velocity much less
than the dynamical (bulk) velocity. This ``cold plasma'' criterion is
well satisfied in X-ray pulsars because the typical inflow speed at the
top of the column is $\sim 0.5\,c$ and the characteristic temperature is
$\sim 10^7\,$K. The resulting X-ray spectrum is characterized by a power
law with photon index $\alpha_{\rm X} > 2$ combined with a low-energy
turnover representing unscattered Planck radiation. The X-ray spectrum
computed using the pure bulk Comptonization model has no high-energy
cutoff, and therefore the photon index must exceed two in order to avoid
an infinite photon energy density. Despite this restriction, the bulk
Comptonization model has successfully reproduced the observed spectra of
several steep-spectrum sources, including X Persei and GX 304-1 (Becker
\& Wolff 2005a,b).

Many of the brightest X-ray pulsars, such as Her X-1 and Cen X-3,
display spectra with photon indices $\alpha_{\rm X} \le 2$ in the $5 -
20\,$keV energy range, combined with high-energy quasi-exponential
cutoffs at $20-30\,$keV (e.g., White et al. 1983). This type of spectral
shape cannot be explained using the pure bulk Comptonization model of
Becker \& Wolff (2005b). The presence of the exponential cutoffs at
high energies, combined with the relatively flat shape at lower
energies, suggests that thermal Comptonization is also playing an
important role in luminous X-ray pulsars by transferring energy from
high- to low-frequency photons via electron scattering. The two steps in
the thermal process (Compton scattering followed by inverse-Compton
scattering) are described mathematically by the Kompaneets (1957)
equation (see also Becker 2003). Although it is clear that the majority
of the photon energization in X-ray pulsars occurs via the first-order
Fermi (i.e., bulk Comptonization process), we find that in bright
sources such as Her X-1, the role of the thermal Comptonization process
must also be considered in order to reproduce the observed spectra.

In this paper, we extend the pure bulk Comptonization model developed by
Becker \& Wolff (2005b) to include both bulk and thermal Comptonization
by incorporating the full Kompaneets operator into the transport
equation used to model the development of the emergent radiation
spectrum. The exact solution to this equation yields the Green's
function for the problem, which represents the contribution to the
observed photon spectrum due to a monochromatic source at a fixed height
in the accretion column. By exploiting the linearity of the mathematical
problem, the Green's function can be used to compute the emergent
spectrum due to any spatial-energetic distribution of photon sources in
the column. We calculate X-ray pulsar spectra by convolving the Green's
function with bremsstrahlung, cyclotron, and blackbody sources, with the
first two distributed throughout the column, and the latter located at
the surface of the thermal mound. The accretion/emission geometry is
illustrated schematically in Figure~1. Seed photons produced inside the
column by various emission mechanisms experience electron scattering as
they diffuse through the column, eventually escaping through the walls
to form the emergent X-ray spectrum. The escaping photons carry away the
kinetic energy of the gas, thereby allowing the plasma to settle onto
the surface of the star.

The remainder of the paper is organized as follows. In \S~2 we briefly
review the nature of the primary radiation transport mechanisms in the
accretion column with a focus on dynamical, thermal, and magnetic
effects. The transport equation governing the formation of the radiation
spectrum is introduced and analyzed in \S~3, and in \S~4 the exact
analytical solution for the Green's function describing the radiation
distribution inside the accretion column is derived. The spectrum of the
radiation escaping through the walls of the accretion column is
developed in \S~5, and the physical constraints for the various model
parameters are considered in \S~6. The nature of the source terms
describing the injection of blackbody, cyclotron, and bremsstrahlung
seed photons into the accretion column is discussed in \S~7. Emergent
X-ray spectra are computed in \S~8, and the results are compared with
the observational data for several luminous X-ray pulsars. The
implications of our work for the production of X-ray spectra in
accretion-powered pulsars are discussed in \S~9.

\section{RADIATIVE PROCESSES}

The dynamics of gas accreting onto the magnetic polar caps of a neutron
star was considered by Basko \& Sunyaev (1976) and Becker (1998). The
formation of the emergent X-ray spectrum in this situation was discussed
by Becker \& Wolff (2005b) based on the geometrical picture illustrated
in Figure~1. Physically, the accretion scenario corresponds to the flow
of a mixture of gas and radiation inside a magnetic ``pipe'' that is
sealed with respect to the gas, but is transparent with respect to the
radiation. The accretion column incorporates a radiation-dominated,
radiative shock located above the stellar surface. Seed photons produced
via a combination of cyclotron, bremsstrahlung, and blackbody radiation
processes are scattered in energy due to collisions with electrons that
are infalling with high speed and also possess a large stochastic
(thermal) velocity component. Blackbody seed photons are produced at the
surface of the dense ``thermal mound'' located at the base of the flow,
where local thermodynamic equilibrium prevails, and cyclotron and
bremsstrahlung seed photons are produced in the optically thin region
above the thermal mound. Hence the surface of the mound represents the
``photosphere'' for photon creation and absorption, and the opacity is
dominated by electron scattering above this point.

\subsection{Magnetic Effects}

The flow of gas in the accretion column of an X-ray pulsar is channeled
by the strong ($B \sim 10^{12}\,$G) magnetic field, and the presence of
this field also has important consequences for the photons propagating
through the plasma. In particular, vacuum polarization leads to
birefringent behavior that gives rise to two linearly polarized normal
modes (Ventura 1979; Nagel 1980; Chanan et al. 1979). The
ordinary mode is polarized with the electric field vector located in the
plane formed by the pulsar magnetic field and the photon propagation
direction. For the extraordinary mode, the electric vector is oriented
perpendicular to this plane. The nature of the photon-electron
scattering process is quite different for the two polarization
modes, and it also depends on whether the photon energy, $\epsilon$,
exceeds the cyclotron energy, $\epsilon_c$, given by
\begin{equation}
\epsilon_c \equiv {e B h \over 2 \pi m_e c}
\approx 11.57 \ B_{12} \ {\rm keV}
\ ,
\label{eq2.1}
\end{equation}
where $B_{12} \equiv B/(10^{12}{\rm G})$ and $c$, $h$, $m_e$, and $e$
represent the speed of light, Planck's constant, and the electron mass
and charge, respectively. Ordinary mode photons interact with electrons
via continuum scattering, governed by the approximate cross section
(Arons et al. 1987)
\begin{equation}
\sigma_{\rm ord}(\epsilon,\varphi) = \sig [\sin^2\varphi
+ k(\epsilon) \, \cos^2\varphi]
\ ,
\label{eq2.2}
\end{equation}
where $\varphi$ is the propagation angle with respect to the magnetic
field, $\sig$ is the Thomson cross section, and
\begin{equation}
k(\epsilon) \equiv \cases{
1 \ , & $\epsilon \ge \epsilon_c$ \ , \cr
(\epsilon / \epsilon_c)^2 \ , & $\epsilon \le \epsilon_c$
\ . \cr
}
\label{eq2.3}
\end{equation}

The scattering cross section for the extraordinary mode photons is more
complex because these photons interact with the electrons via both
continuum and resonant processes. In this case the total cross section
can be approximated using (Arons et al. 1987)
\begin{equation}
\sigma_{\rm ext}(\epsilon,\varphi) = \sig k(\epsilon)
+ \sigma_l \, \phi_l(\epsilon,\epsilon_c,\varphi)
\ ,
\label{eq2.4}
\end{equation}
where $\phi_l$ is the unity normalized line profile function (which is
resonant at the cyclotron energy $\epsilon_c$), and
\begin{equation}
\sigma_l = 1.9 \times 10^4 \, \sig \, B_{12}^{-1}
\ .
\label{eq2.5}
\end{equation}
Equation~(\ref{eq2.4}) indicates that the extraordinary mode photons
experience a substantial enhancement in the scattering cross section
close to the cyclotron energy, which reflects their ability to cause
radiative excitation of electrons from the ground state to the first
excited Landau level (Ventura 1979). The excitation process is almost
always followed by a radiative deexcitation, and therefore cyclotron
absorption can be viewed as a form of resonant scattering (Nagel 1980).
The resonant nature of the cyclotron interaction gives rise to a strong
high-energy absorption feature in many observed X-ray pulsar spectra.

The energy and angular dependences of the electron scattering cross
section are quite different for the two polarization modes. For
radiation propagating perpendicular to the magnetic field with energy
$\epsilon < \epsilon_c$, photons in the ordinary mode will see a cross
section that is essentially Thomson, whereas the extraordinary mode
photons will experience a cross section that is reduced from the Thomson
value by the ratio $(\epsilon/\epsilon_c)^2$. On the other hand, in the
limit of parallel or antiparallel propagation relative to the field
direction, both the ordinary and extraordinary mode cross sections are
reduced by the factor $(\epsilon/\epsilon_c)^2$ relative to the Thomson
value if $\epsilon < \epsilon_c$. In practice, the two modes communicate
via mode conversion, which occurs at the continuum scattering rate
(Arons et al. 1987). Hence if $\epsilon$ is sufficiently below the
cyclotron energy $\epsilon_c$ so that the resonant contribution to the
cross section in equation~(\ref{eq2.4}) is negligible, then the
scattering of photons propagating perpendicular to the magnetic field is
dominated by the ordinary mode cross section, $\sigma_{\rm ord} = \sig$,
because this yields the smallest mean free path.

Due to the importance of radiation pressure and the complexity of the
electron scattering cross sections, the dynamical structure of the flow
is closely tied to the spatial and energetic distribution of the
radiation. The coupled radiation-hydrodynamical problem is so complex
and nonlinear that it is essentially intractable. In order to make
progress a set of simplifying assumptions must be adopted. In
particular, a detailed consideration of the angular and energy
dependences of the electron scattering cross sections for the two
polarization modes is beyond the scope of the present paper. We shall
therefore follow Wang \& Frank (1981) and Becker (1998) by treating the
directional dependence of the electron scattering in an approximate way
in terms of the constant, energy- and mode-averaged cross sections
$\sigpar$ and $\sigperp$ describing respectively the scattering of
photons propagating either parallel or perpendicular to the magnetic
field. For the magnetic field strengths $B \sim 10^{12-13}\,$G and
electron temperatures $T_e \sim 10^{6-7}\,$K typical of luminous X-ray
pulsars, the mean photon energy, $\bar\epsilon$, is well below
$\epsilon_c$. In this case the mean scattering cross section for photons
propagating parallel to the field can be approximated by writing (see
eqs.~[\ref{eq2.2}] and [\ref{eq2.4}])
\begin{equation}
\sigpar \approx \sig \left(\bar\epsilon \over \epsilon_c\right)^2
\ ,
\label{eq2.6}
\end{equation}
and the mean scattering cross section for photons propagating perpendicular
to the field (which is dominated by the ordinary mode) is given by
\begin{equation}
\sigperp \approx \sig
\ .
\label{eq2.7}
\end{equation}
Note the substantial decrease in the opacity of the gas along the
magnetic field indicated by equation~(\ref{eq2.6}), which is consistent
with results presented by Canuto et al. (1971).

In our calculations, the perpendicular scattering cross section
$\sigperp$ is set equal to the Thomson value using
equation~(\ref{eq2.7}). However, the value of the parallel scattering
cross section $\sigpar$ is more problematic since the mean energy
$\bar\epsilon$ cannot be calculated until the radiative transfer problem
is solved. In practice, the value of $\sigpar$ is calculated using a
dynamical relationship based on the flow of radiation-dominated gas in a
pulsar accretion column, as discussed in \S~6, where we also examine the
self-consistency of our results.

\subsection{Radiation-Dominated Flow}

Radiation pressure governs the dynamical structure of the accretion
flows in bright pulsars when the X-ray luminosity $\xlum$ satisfies (Becker
1998; Basko \& Sunyaev 1976)
\begin{equation}
\xlum \sim L_{\rm crit} \equiv
{2.72 \times 10^{37} \sig \over
\sqrt{\sigperp\sigpar}}
\left(M_* \over \msun \right)
\left(\colrad \over R_*\right)
\ {\rm \ ergs \ s}^{-1}
\ ,
\label{eq2.8}
\end{equation}
where $\colrad$ is the polar cap radius, $\starmass$ and $\starad$
denote the stellar mass and radius, respectively, and $\xlum$ is
related to the accretion rate $\dot M$ via
\begin{equation}
\xlum = {G M_* \dot M \over R_*}
\ .
\label{eq2.8b}
\end{equation}
When the luminosity of the system is comparable to $L_{\rm crit}$, the
radiation flux in the column is super-Eddington and therefore the
radiation pressure greatly exceeds the gas pressure (Becker 1998). In
this situation the gas passes through a radiation-dominated shock on its
way to the stellar surface, and the kinetic energy of the particles is
carried away by the high-energy radiation that escapes from the column.
The strong gradient of the radiation pressure decelerates the material
to rest at the surface of the star. The observation of many X-ray
pulsars with $L_{\rm X} \sim 10^{36-38} \, {\rm ergs \, s}^{-1}$ implies
the presence of radiation-dominated shocks close to the stellar surfaces
in these systems (White et al. 1983; White et al. 1995). Note that
radiation-dominated shocks are {\it continuous} velocity transitions,
with an overall thickness of a few Thomson scattering lengths, unlike
traditional (discontinuous) gas-dominated shocks (Blandford \& Payne
1981b). Despite the fact that the luminosities of the brightest X-ray
pulsars exceed the Eddington limit for a neutron star, the material in
the accretion column is decelerated to rest at the stellar surface,
rather than being blown away, because the scattering cross section
$\sigpar$ for photons propagating parallel to the magnetic field is
generally much smaller than the Thomson value.

Becker \& Wolff (2005b) considered the effects of pure bulk
Comptonization on the emergent spectrum in an X-ray pulsar based on the
exact velocity profile derived by Becker (1998) and Basko \& Sunyaev
(1976), which describes the accretion of gas onto a neutron star through
a standing, radiation-dominated shock. In the present paper, our goal is
to extend the model to include the effects of thermal Comptonization,
which, along with cyclotron absorption, is expected to produce a
steepening of the spectrum at high energies, as observed in many
luminous X-ray pulsars. The thermal process will also cause a flattening
of the spectrum at lower energies, relative to the spectrum resulting
from pure bulk Comptonization, due to the redistribution of energy from
high- to medium-energy photons via electron recoil. Since there are many
physical effects involved in the scenario considered here, this is
clearly a rather complex problem. In order to render the problem
mathematically tenable, we will utilize an approximate velocity profile
that agrees reasonably well with the exact profile derived by Becker
(1998) and used by Becker \& Wolff (2005a,b) in their study of pure bulk
Comptonization in X-ray pulsars. The validity of this approximation will
be carefully examined by comparing the spectrum obtained using the
approximate velocity profile with that computed using the pure bulk
Comptonization model of Becker \& Wolff in the limit of zero electron
temperature, in which case both models should agree.

\subsection{Thermal Equilibration}

Another important question concerns the nature of the energy
distribution in the accreting plasma. In an X-ray pulsar accretion
column, the electrons will have an anisotropic energy distribution,
described by a one-dimensional Maxwellian with a relatively high
temperature ($T_e \sim 10^7\,$K) along the direction of the magnetic
field, and an essentially monoenergetic distribution in the
perpendicular direction (e.g., Arons et al. 1987). Most of the electrons
reside in the lowest Landau state (the ground state), and these
particles possess no gyrational motion. However, the electrons in the
first excited Landau level have energy $\epsilon_c \sim 20-40\,$keV (see
eq.~[\ref{eq2.1}]), which is much larger than the typical thermal energy
in the parallel direction. The protons, which are not as strongly
effected by the field, have a three-dimensional thermal distribution.
Collisions with high speed protons can cause the electrons to be excited
to higher Landau levels, followed rapidly by radiative deexcitation.
This process essentially converts proton kinetic energy into radiation
energy via the production of cyclotron photons, which cools the plasma.
The Comptonization of these seed photons leads to further cooling before
the radiation escapes through the walls of the accretion column in the
form of X-rays.

In principle, the temperature of the ions may depart significantly from
the electron temperature, depending on the ratio of the Coulomb coupling
rate to the dynamical timescale for accretion onto the stellar surface.
We can determine if the gas in the X-ray pulsar accretion column has a
two-temperature structure by comparing the electron-proton equilibration
timescale with the dynamical and radiative cooling timescales. Based on
equation~(103) from Arons et al. (1987), we can approximate the
electron-ion coupling timescale using
\begin{equation}
t_{ei} \sim 10^{-11} \, \left(\rho \over 0.01 \, {\rm g\,cm}^{-3}\right)^{-1}
\left(T_e \over 10^7 \, {\rm K}\right)^{3/2} \ {\rm s}
\ ,
\label{eq2.9}
\end{equation}
where $\rho$ is the density of the accreting gas. In the bright sources
of interest here, we typically find that $T_e \sim 10^{6-7}\,$K and
$\rho \sim 10^{-4} - 10^{-2} \, {\rm g \, cm}^{-3}$ within the radiating
portion of the accretion column, giving $t_{ei} \sim 10^{-9} - 10^{-13}\,$s.

The opacity of the gas in the accretion column of a bright X-ray pulsar
is dominated by electron scattering, and therefore inverse-Compton
scattering of soft photons by the hot electrons is the primary cooling
mechanism. The rate of change of the photon energy density due to
inverse-Compton scattering is given by (see eq.~[7.22] from Rybicki \&
Lightman 1979)
\begin{equation}
{dU \over dt} \Bigg|_{\rm IC} = n_e \, \sigbar \, c \ {4 k T_e \over m_e c^2}
\ U
\ ,
\label{eq2.10}
\end{equation}
where $n_e=\rho/m_p$ is the electron number density for pure,
fully-ionized hydrogen, $m_p$ is the proton mass, $U$ denotes the photon
energy density, and $\sigbar$ represents the angle-averaged electron
scattering cross section, which depends on the energy and angular
distribution of the radiation field. In general, we expect to find that
$\sigpar < \sigbar < \sigperp$, where $\sigpar$ and $\sigperp$ are the
scattering cross sections for photons propagating either parallel or
perpendicular to the magnetic field (see eqs.~[\ref{eq2.6}] and
[\ref{eq2.7}]).

The photon energy density $U$ can be estimated in terms of the X-ray
luminosity $\xlum$ and the column radius $r_0$ using
\begin{equation}
U \sim {\xlum \over \pi r_0^2 \, c}
\ .
\label{eq2.11}
\end{equation}
By combining equation~(\ref{eq2.10}) and (\ref{eq2.11}), we find that the
inverse-Compton cooling timescale for the electrons is given by
\begin{equation}
t_{\rm IC} \equiv {(3/2) n_e k T_e \over dU/dt}
\sim 10^{-6} \, \left(\sigbar \over 10^{-3} \, \sig\right)^{-1}
\left(r_0 \over 1\,{\rm km}\right)^2
\left(\xlum \over 10^{37}\,{\rm ergs \ s}^{-1}\right)^{-1} \ {\rm s}
\ .
\label{eq2.12}
\end{equation}
In the X-ray pulsar application, we typically obtain (see \S~6) $\sigbar
/\sig \sim 10^{-4}-10^{-3}$, $r_0 \sim 0.1-1\,$km, and $\xlum \lapprox
10^{37}\, {\rm ergs \ s}^{-1}$, and therefore $t_{\rm IC} \gapprox
10^{-8}\,$s. Next we recall that the characteristic dynamical
(free-fall) timescale for neutron star accretion is given by
\begin{equation}
t_{\rm dyn} \equiv \left(R_*^3 \over 2 \, G M_*\right)^{1/2}
\sim 10^{-4} \left(R_* \over 10\,{\rm km}\right)^{3/2}
\left(M_* \over \msun\right)^{-1/2} \ {\rm s}
\ .
\label{eq2.13}
\end{equation}
It is apparent that the dynamical timescale $t_{\rm dyn}$ and the
inverse-Compton cooling timescale $t_{\rm IC}$ each exceed the
electron-ion equilibration timescale $t_{ei}$ by several orders of
magnitude, and therefore we conclude that the electron and ion
temperatures are essentially equal in the sources of interest here.

\begin{figure}[t]
\begin{center}
\hskip-0.4truein
\epsfig{file=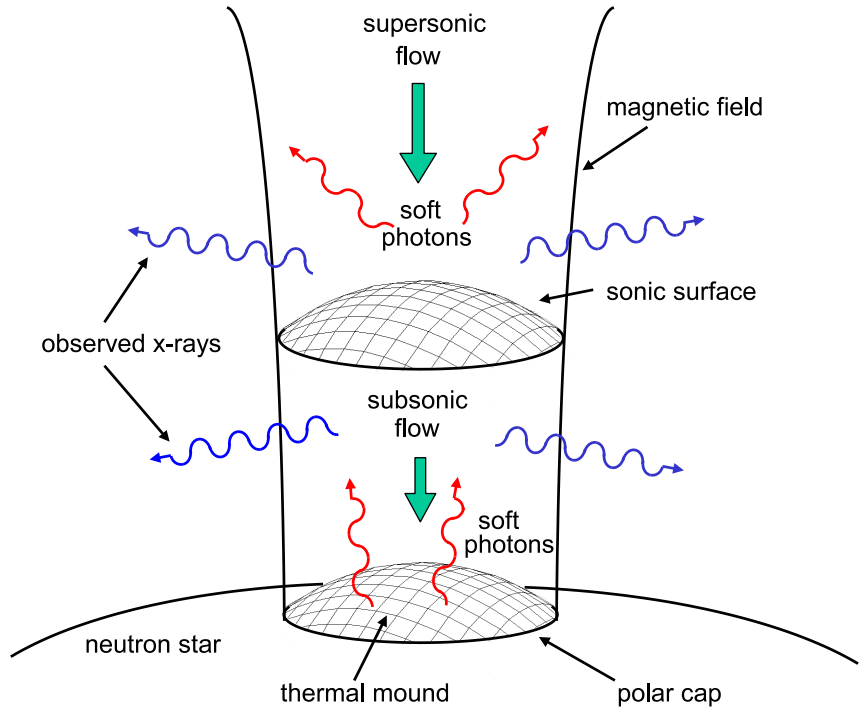,height=13.0cm,angle=0}
\end{center}
\vskip-1.0truein
\caption{Schematic depiction of gas accreting onto the magnetic polar
cap of a neutron star. Seed photons are created throughout the column
via bremsstrahlung and cyclotron emission, and additional blackbody seed
photons are emitted from the surface of the thermal mound, near the base
of the column.}
\end{figure}

\subsection{Thermal and Bulk Comptonization}

The strong compression that occurs as the plasma crosses the radiative
shock renders it an ideal site for the bulk Comptonization of seed
photons produced via bremsstrahlung, cyclotron, or blackbody emission.
In the bulk Comptonization process, particles experience a mean energy
gain if the scattering centers they collide with are involved in a
converging flow (e.g., Laurent \& Titarchuk 1999; Turolla, Zane, \&
Titarchuk 2002). By contrast, in the thermal Comptonizaton process,
particles gain energy due to the {\it stochastic} motions of the
scattering centers via the second-order Fermi mechanism (e.g., Sunyaev
\& Titarchuk 1980; Becker 2003). In the X-ray pulsar application, the
scattering centers are infalling electrons, and the energized
``particles'' are photons. Since the inflow speed of the electrons in an
X-ray pulsar accretion column is much larger than their thermal
velocity, bulk Comptonization dominates over the stochastic process
except at the highest photon energies (Titarchuk, Mastichiadis, \&
Kylafis 1996).

The fundamental character of the Green's function describing both
thermal and bulk Comptonization was studied by Titarchuk \& Zannias
(1998) for the case of accretion onto a black hole. These authors
established that the Green's function can be approximated using a broken
power-law form with a central peak between the high- and low-energy
portions of the spectrum, for either type of Comptonization.
Furthermore, they concluded that bulk Comptonization dominates over the
thermal process if the electron temperature $T_e \lapprox 10^7\,$K. The
theory developed in the present paper represents an extension of the
same idea to neutron star accretion, resulting in the exact Green's
function for the X-ray pulsar spectral formation process. In agreement
with Titarchuk \& Zannias (1998), we find that the bulk Comptonization
process dominates the energy exchange between the electrons and the
photons for the temperature range relevant in X-ray pulsars. However, we
also establish that thermal Comptonization nonetheless plays a central
role in forming the characteristic {\it spectral shapes} noted in the
bright sources. For example, Compton scattering of the high-energy
photons is crucial for producing the observed quasi-exponential
turnovers at high energies as a result of electron recoil, which is part
of the thermal Comptonization process represented by the Kompaneets
operator discussed in \S~3. The subsequent inverse-Compton scattering of
soft photons by the recoiling electrons redistributes the energy to
lower frequency photons, helping to flatten the spectra as observed.

Lyubarskii \& Sunyaev (1982) analyzed the effect of bulk and thermal
Comptonization in the context of neutron star accretion using the same
dynamical model for the flow considered here. However, they did not
include the effect of photon escape from the accretion column, which
renders their model inapplicable for the modeling of spectral formation
in X-ray pulsars. In this paper we extend the work of Lyubarskii \&
Sunyaev by including the crucial effect of photon escape. Poutanen \&
Gierli\'nski (2003) computed X-ray pulsar spectra based on the thermal
Comptonization of soft radiation in a hot layer above the magnetic pole,
but their model did not include a complete treatment of the bulk
process, which is of central importance in X-ray pulsars. The new theory
presented here therefore represents the first exact, quantitative
analysis of the role of bulk and thermal Comptonization in the X-ray
pulsar spectral formation process.

\section{FORMATION OF THE RADIATION SPECTRUM}

We follow the approach of Basko \& Sunyaev (1975, 1976) and Becker \&
Wolff (2005b), and assume that the upstream flow is composed of pure,
fully-ionized hydrogen moving at a highly supersonic speed, which is the
standard scenario for accretion-powered X-ray pulsars. Our transport
model employs a cylindrical, plane-parallel geometry, and therefore the
velocity, density, and pressure are functions of the distance above the
stellar surface, but they are all constant across the column at a given
height (see Fig.~1). In the region above the thermal mound in a luminous
X-ray pulsar, the gas is radiation-dominated, and the photons interact
with the matter primarily via electron scattering, which controls both
the spatial transport and the energization of the radiation (Arons et
al. 1987). ``Seed'' photons injected into the flow are unable to diffuse
very far up into the accreting gas due to the extremely high speed of
the inflow. Most of the photons therefore escape through the walls of
the column within a few scattering lengths of the mound, forming a
``fan'' type beam pattern, as expected for accretion-powered X-ray
pulsars (e.g., Harding 1994, 2003). We are interested in obtaining the
steady-state, polarization mode-averaged photon distribution function,
$f(z,\epsilon)$, measured at altitude $z$ and energy $\epsilon$ inside
the column resulting from the reprocessing of blackbody, cyclotron, and
bremsstrahlung seed photons. The normalization of $f$ is defined so that
$\epsilon^2 f(z,\epsilon) \, d\epsilon$ gives the number density of
photons in the energy range between $\epsilon$ and $\epsilon+d\epsilon$,
and therefore $f$ is related to the occupation number distribution $\bar
n$ via $f = 8 \pi \, \bar n / (c^3 h^3)$.

\subsection{Transport Equation}

In the cylindrical, plane-parallel geometry employed here, the photon
distribution $f(z,\epsilon)$ satisfies the transport equation (e.g.,
Becker \& Begelman 1986; Blandford \& Payne 1981a; Becker 2003)
\begin{eqnarray}
{\partial f \over \partial t} + v \, {\partial f \over \partial z}
&=& {dv \over d z}\,{\epsilon\over 3} \,
{\partial f\over\partial\epsilon}
+ {\partial\over\partial z}
\left({c\over 3 n_e \sigpar}\,{\partial f\over\partial z}\right)
- {f \over t_{\rm esc}}
\nonumber
\\
&+& {n_e \sigbar c \over m_e c^2} {1 \over\epsilon^2}
{\partial\over\partial\epsilon}\left[\epsilon^4\left(f
+ k T_e \, {\partial f\over\partial\epsilon}\right)\right]
+ {Q(z,\epsilon) \over \pi r_0^2}
\ ,
\label{eq3.1}
\end{eqnarray}
where $z$ is the distance from the stellar surface along the column
axis, $v < 0$ is the inflow velocity, $Q$ denotes the photon source
distribution, and $t_{\rm esc}$ represents the mean time photons spend
in the plasma before diffusing through the walls of the column. The
source function $Q$ is normalized so that $\epsilon^2 \, Q(z,\epsilon)
\, d\epsilon \, dz$ gives the number of seed photons injected per unit
time between $z$ and $z+dz$ with energy between $\epsilon$ and
$\epsilon+d\epsilon$. The left-hand side of equation~(\ref{eq3.1})
denotes the comoving time derivative of the radiation distribution $f$,
and the terms on the right-hand side represent first-order Fermi
energization (``bulk Comptonization''), spatial diffusion along the
column axis, photon escape, thermal Comptonization, and photon
injection, respectively. We are interested here in the steady-state
version of equation~(\ref{eq3.1}) with $\partial f/\partial t=0$. The
transport equation employed here is similar to the one analyzed by
Becker \& Wolff (2005b), except that thermal Comptonization has now been
included via the appearance of the Kompaneets (1957) operator, and the
source term $Q$ has been generalized to treat the production of seed
radiation throughout the column, rather than focusing exclusively on the
injection of blackbody photons at the surface of the thermal mound.
Equation~(\ref{eq3.1}) does not include the term used by Becker \& Wolff
(2005b) to describe absorption at the thermal mound surface because this
effect is negligible when bremsstrahlung and cyclotron emission are
included in the model, as discussed in \S~8. The total photon number and
energy densities associated with the radiation distribution $f$ are
given, respectively, by
\begin{equation}
n(z) = \int_0^\infty \epsilon^2 \,
f(z,\epsilon) \, d\epsilon
\ , \ \ \ \ 
U(z) = \int_0^\infty \epsilon^3 \,
f(z,\epsilon) \, d\epsilon
\ .
\label{eq3.2}
\end{equation}

Following Becker (1998) and Becker \& Wolff (2005b), we compute the
mean escape time using the diffusive prescription
\begin{equation}
t_{\rm esc}(z) = {r_0 \, \tauperp \over c} \ , \ \ \ \ \ 
\tauperp(z) = n_e \, \sigperp \, \colrad
\ ,
\label{eq3.3}
\end{equation}
where $\tauperp$ represents the perpendicular scattering optical
thickness of the cylindrical accretion column, and $\tauperp$ and
$t_{\rm esc}$ are each functions of $z$ through their dependence on the
electron number density $n_e$. Becker (1998) confirmed that the
diffusion approximation employed in equation~(\ref{eq3.3}) is valid
since $\tauperp > 1$ for typical X-ray pulsar parameters.
Equation~(\ref{eq3.3}) for the mean escape timescale can be rewritten as
\begin{equation}
t_{\rm esc}(z) = {\dot M \, \sigperp \over \pi m_p \, c \, |v|}
\ ,
\label{eq3.4}
\end{equation}
where
\begin{equation}
\dot M \equiv \pi r_0^2 \, \rho \, |v| = {\rm constant}
\ ,
\label{eq3.5}
\end{equation}
and $\rho=m_p\,n_e$. Since the escape timescale is inversely
proportional to the flow velocity, the column becomes completely opaque
at the surface of the neutron star due to the divergence of the electron
number density there. The relationship between the escape-probability
formalism employed here and the physical distribution of radiation
inside the accretion column was discussed in detail by Becker \& Wolff
(2005b) in their \S~6.3.

In our approach to solving equation~(\ref{eq3.1}) for the spectrum
$f(z,\epsilon)$ inside the accretion column, we shall first obtain
the Green's function, $\green(z_0,z,\epsilon_0,\epsilon)$, which is the
radiation distribution at location $z$ and energy $\epsilon$ resulting
from the injection of $\dot N_0$ photons per second with energy
$\epsilon_0$ from a monochromatic source at location $z_0$. The
determination of the Green's function is a useful intermediate step in
the process because it provides us with fundamental physical insight
into the spectral redistribution process, and it also allows us to
calculate the particular solution for the spectrum $f$ associated with
an arbitrary photon source $Q(z,\epsilon)$ using the integral
convolution (Becker 2003)
\begin{equation}
f(z,\epsilon) = \int_0^\infty\int_0^\infty
{\green(z_0,z,\epsilon_0,\epsilon) \over \dot N_0} \ \epsilon_0^2
\, Q(z_0,\epsilon_0) \, d\epsilon_0 \, dz_0
\ .
\label{eq3.6}
\end{equation}
The technical approach used to solve for the Green's function, carried
out in \S~4, involves the derivation of eigenvalues and associated
eigenfunctions based on the set of spatial boundary conditions for the
problem (see, e.g., Blandford \& Payne 1981b; Payne \& Blandford 1981;
Schneider \& Kirk 1987; Colpi 1988).

The steady-state transport equation governing the Green's function
$\green$ is (cf. eq.~[\ref{eq3.1}])
\begin{eqnarray}
v \, {\partial \green \over \partial z}
&=& {dv \over d z}\,{\epsilon\over 3} \,
{\partial \green\over\partial\epsilon}
+ {\partial\over\partial z}
\left({c\over 3 n_e \sigpar}\,{\partial \green\over\partial z}\right)
- {\green \over t_{\rm esc}}
\nonumber
\\
&+& {n_e \sigbar c \over m_e c^2} {1 \over\epsilon^2}
{\partial\over\partial\epsilon}\left[\epsilon^4\left(\green
+ k T_e \, {\partial \green\over\partial\epsilon}\right)\right]
+ {\dot N_0 \, \delta(\epsilon-\epsilon_0) \, \delta(z-z_0)
\over \pi r_0^2 \epsilon_0^2}
\ ,
\label{eq3.7}
\end{eqnarray}
and the associated radiation number and energy densities are given
by (cf. eqs.~[\ref{eq3.2}])
\begin{equation}
\ngreen(z) \equiv \int_0^\infty \epsilon^2 \,
\green(z_0,z,\epsilon_0,\epsilon) \, d\epsilon \ , \ \ \ \ \ 
\ugreen(z) \equiv \int_0^\infty \epsilon^3 \,
\green(z_0,z,\epsilon_0,\epsilon) \, d\epsilon
\ .
\label{eq3.8}
\end{equation}
Following Lyubarskii \& Sunyaev (1982), we will assume that the electron
temperature $T_e$ has a constant value, which is physically reasonable
since most of the Comptonization occurs in a relatively compact region
above the thermal mound. However, it is important to confirm the
validity of this assumpton using a detailed numerical model that
incorporates a self-consistent calculation of the temperature
distribution, which we plan to develop in future work. Under the
assumption of a constant electron temperature, it is convenient to work
in terms of the dimensionless energy variable $\chi$, defined by
\begin{equation}
\chi(\epsilon) \equiv {\epsilon \over kT_e}
\ .
\label{eq3.9}
\end{equation}
We can make further progress by transforming the spatial variable from
$z$ to the scattering optical depth parallel to the magnetic field,
$\tau$, which is related to $z$ via
\begin{equation}
d\tau = n_e(z) \, \sigpar \, dz
\ , \ \ \ \
\tau(z) = \int_0^z n_e(z') \, \sigpar \, dz'
\ ,
\label{eq3.10}
\end{equation}
so that $z$ and $\tau$ both vanish at the stellar surface.

Making the change of variable from $z$ to $\tau$ in
equation~(\ref{eq3.7}), we find after some algebra that the transport
equation for the Green's function can be written in the form
\begin{eqnarray}
{v \over c} \, {\partial \green \over \partial \tau}
&=& {1 \over c}\,{dv \over d\tau}\,{\chi\over 3} \,
{\partial \green\over\partial\chi}
+ {1 \over 3}\,{\partial^2\green\over\partial\tau^2}
- {\xi^2\,v^2 \over c^2} \, \green
\nonumber
\\
&+& {\sigbar \over \sigpar} \, {kT_e \over m_e c^2}
{1 \over\chi^2}{\partial\over\partial\chi}\left[\chi^4\left(\green
+ {\partial\green\over\partial\chi}\right)\right]
+ {\dot N_0 \, \delta(\chi-\chi_0) \, \delta(\tau-\tau_0)
\over \pi r_0^2 c k T_e \epsilon_0^2}
\ ,
\label{eq3.11}
\end{eqnarray}
where $\tau_0 \equiv \tau(z_0)$, $\chi_0 \equiv \chi(\epsilon_0)$, and
we have introduced the dimensionless parameter
\begin{equation}
\xi \equiv {\pi r_0 m_p c \over \dot M (\sigpar \sigperp)^{1/2}}
\ ,
\label{eq3.12}
\end{equation}
which determines the importance of the escape of photons from the
accretion column. Becker (1998) derived the exact solution for the flow
velocity profile in a radiation-dominated pulsar accretion column, and
demonstrated that the condition $\xi = 2/\sqrt{3}$ must be satisfied in
order to ensure that the flow comes to rest at the stellar surface. This
condition represents a balance between the characteristic timescale for
photon escape and the dynamical (accretion) timescale, as discussed in
\S~6. Physically, this balance reflects the requirement that the kinetic
energy of the flow must be radiated away through the column walls in the
same amount of time required for the gas to settle onto the star. Note
that we can write the Green's function as either $\green(z_0,z,
\epsilon_0,\epsilon)$ or $\green(\tau_0,\tau,\chi_0,\chi)$ since the
variables $(z,\epsilon)$ and $(\chi,\tau)$ are interchangeable via
equations~(\ref{eq3.9}) and (\ref{eq3.10}).

\subsection{Separability}

Lyubarskii \& Sunyaev (1982) demonstrated that when $\chi \ne \chi_0$,
the transport equation~(\ref{eq3.11}) is separable in energy and space
if the velocity profile has the particular form
\begin{equation}
v(\tau) = - \alpha \, c \, \tau
\ ,
\label{eq3.13}
\end{equation}
where $\alpha$ is a positive constant. By combining equations~(\ref{eq3.5}),
(\ref{eq3.10}), (\ref{eq3.12}), and (\ref{eq3.13}), we can express $\tau$
as an explicit function of the altitude $z$, obtaining
\begin{equation}
\tau(z) = \left(\sigpar \over \sigperp\right)^{1/4}
\left(2 \, z \over \alpha \, \xi \, r_0\right)^{1/2}
\ .
\label{eq3.14}
\end{equation}
Using this result to substitute for $\tau$ in equation~(\ref{eq3.13}),
we note that the velocity profile required for separability is related
to $z$ via
\begin{equation}
v(z) = - \left(\sigpar \over \sigperp\right)^{1/4}
\left(2 \, \alpha z \over \xi \, r_0\right)^{1/2} c
\ .
\label{eq3.15}
\end{equation}
Although this profile describes a flow that stagnates at the stellar
surface ($\tau=0$, $z=0$) as required, the details of the velocity
variation deviate somewhat from the exact solution for the velocity
profile in a radiation-dominated pulsar accretion column derived by
Becker (1998) and Basko \& Sunyaev (1976), which can be stated in terms
of the altitude $z$ using
\begin{equation}
\vex(z) = - \vff \left[1-\left(7 \over 3\right)^{-z/\zst}\right]
\ ,
\label{eq3.16}
\end{equation}
where the free-fall velocity from infinity onto the stellar surface,
$\vff$, and the altitude at the sonic point, $\zst$, are given by
\begin{equation}
\vff \equiv \left(2 \, G M_* \over R_*\right)^{1/2}
\ , \ \ \ \
\zst \equiv {r_0 \over 2\sqrt{3}} \left(\sigperp\over\sigpar\right)^{1/2}
\ln\left(7 \over 3\right)
\ .
\label{eq3.17}
\end{equation}

In order to make further mathematical progress in the computation of the
radiation spectrum emitted by the accretion column, we must utilize the
approximate (separable) velocity profile given by
equation~(\ref{eq3.15}). However, before doing so, we must ensure that
the approximate velocity profile agrees reasonably well with the exact
profile (eq.~[\ref{eq3.16}]) in the lower portion of the accretion
column, where most of the spectral formation occurs. We can accomplish
this by setting the two profiles equal to each other at the sonic point
($z=\zst$), and using this condition to solve for the constant $\alpha$,
which yields
\begin{equation}
\alpha = {32 \, \sqrt{3} \over 49 \, \ln(7/3)}
\ {GM_* \, \xi \over R_* \, c^2}
\ ,
\label{eq3.18}
\end{equation}
or, equivalently,
\begin{equation}
\alpha = 0.20
\left(M_* \over \msun\right)
\left(R_* \over 10\,{\rm km}\right)^{-1} \xi
\ .
\label{eq3.19}
\end{equation}
This relation allows us to compute $\alpha$ as a function of $\xi$ for
given values of the stellar mass $M_*$ and radius $R_*$. In
radiation-dominated pulsar accretion columns, $\xi=2/\sqrt{3}$, and
therefore $\alpha$ is of order unity (Becker 1998).

\begin{figure}[t]
\begin{center}
\epsfig{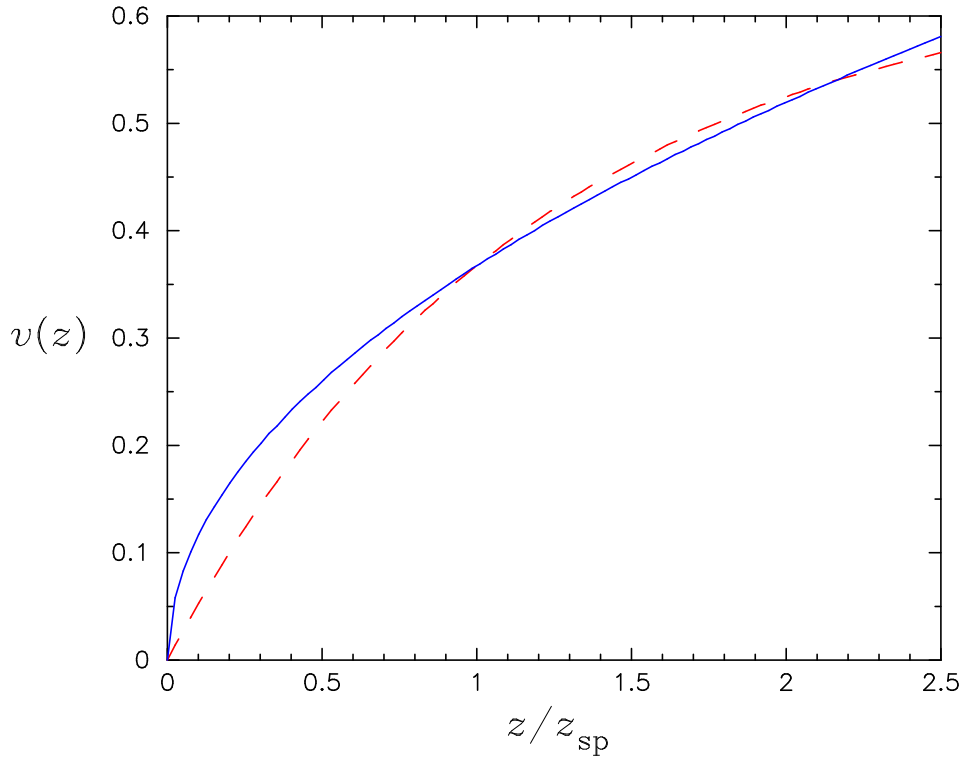}
\end{center}
\caption{Comparison of the approximate ({\it solid line}) and exact
({\it dashed line}) velocity profiles evaluated using
eqs.~(\ref{eq3.15}) and (\ref{eq3.16}), respectively, in units of $c$.
The constant $\alpha$ has been computed using eq.~(\ref{eq3.18}), which
ensures that the two velocities agree at the sonic point, $z=\zst$.}
\end{figure}

The detailed shapes of the approximate (eq.~[\ref{eq3.15}]) and exact
(eq.~[\ref{eq3.16}]) velocity profiles as functions of $z$ are compared
in Figure~2. In general, the two functions agree fairly well, although
the approximate profile overestimates the correct velocity close to the
stellar surface. Nevertheless, since the overall compression associated
with the approximate profile is correct, we expect that the net effect
of bulk Comptonization will be accurately modeled using the approximate
profile. We will therefore adopt the form for the velocity profile given
in terms of $\tau$ by equation~(\ref{eq3.13}) or in terms of $z$ by
equation~(\ref{eq3.15}), with $\alpha$ computed using
equation~(\ref{eq3.19}). The validity of this approach will be
carefully examined in \S~6 by comparing the results obtained for the
Comptonization of radiation in a ``cold'' accretion flow ($T_e \to 0$)
described by the approximate velocity profile with those computed using
the model of Becker \& Wolff (2005b), who treated the Comptonization
of radiation in a cold accretion column, described by the exact velocity
profile.

\section{EXACT SOLUTION FOR THE GREEN'S FUNCTION}

Adopting the velocity profile given by equation~(\ref{eq3.13}), we find
that the transport equation~(\ref{eq3.11}) for the Green's function can
be reorganized to obtain
\begin{eqnarray}
{\alpha \, \chi\over 3} \,
{\partial \green \over \partial \chi}
- {\sigbar \over \sigpar} \, {kT_e \over m_e c^2}
{1 \over\chi^2}{\partial\over\partial\chi}\left[\chi^4\left(\green
+ {\partial\green\over\partial\chi}\right)\right]
&=& {1 \over 3}\,{\partial^2\green\over\partial\tau^2}
+ \alpha \, \tau \, {\partial \green \over \partial \tau}
- \xi^2 \alpha^2\,\tau^2 \, \green
\nonumber
\\
&+& {\dot N_0 \, \delta(\chi-\chi_0) \, \delta(\tau-\tau_0)
\over \pi r_0^2 c k T_e \epsilon_0^2}
\ .
\label{eq4.1}
\end{eqnarray}
Lyubarskii \& Sunyaev (1982) analyzed this equation for the special
case $\xi=0$, which corresponds to the neglect of the escape of photons
from the accretion column. Their results do not describe the formation of
the emergent spectrum in an accretion-powered X-ray pulsar since photon
escape is a critical part of that process. We therefore extend the results
of Lyubarskii \& Sunyaev below by deriving the closed-form solution for
the Green's function in the X-ray pulsar problem for the general case
with $\xi > 0$.

When $\chi \ne \chi_0$, the $\delta$-function in the transport
equation~(\ref{eq4.1}) makes no contribution, and therefore the
differential equation is linear and homogeneous. The transport equation
can then be separated in energy and space using the functions
\begin{equation}
f_\lambda(\tau,\chi) \equiv g(\lambda,\tau) \ h(\lambda,\chi) \ ,
\label{eq4.2}
\end{equation}
where $\lambda$ is the separation constant. We find that the spatial
and energy functions, $g$ and $h$, respectively, satisfy the differential
equations
\begin{equation}
{1 \over 3}\,{d^2 g \over d\tau^2}
+ \alpha \, \tau \, {d g \over d\tau}
+ \left({\alpha \lambda \over 3}
- \xi^2 \alpha^2 \tau^2\right) g = 0
\ ,
\label{eq4.3}
\end{equation}
and
\begin{equation}
{1 \over \chi^2}{d\over d\chi}\left[\chi^4
\left(h + {d h\over d\chi}\right)\right]
- \deltapar \, \chi \,
{d h \over d\chi}
- \deltapar \, \lambda \, h = 0
\ ,
\label{eq4.4}
\end{equation}
where the parameter $\deltapar$ is defined by
\begin{equation}
\deltapar \equiv {\alpha \over 3} \,
{\sigpar \over \sigbar} \, {m_e c^2 \over kT_e}
\ .
\label{eq4.5}
\end{equation}
This is equivalent to the quantity $\delta$ introduced by Lyubarskii
\& Sunyaev (1982) if we set $\sigbar=\sigpar$, since these authors did
not include any angle dependence in the electron scattering cross section.

\subsection{Eigenvalues and Spatial Eigenfunctions}

In order to obtain the solution for the spatial separation function $g$,
we must first consider the boundary conditions that $g$ must satisfy. In
the downstream region, as the gas approaches the stellar surface, we
expect the advective and diffusive components of the radiation flux to
vanish due to the divergence of the electron density. The advective flux
is indeed negligible at the stellar surface since $v \to 0$ as $\tau \to
0$ (see eq.~[\ref{eq3.13}]). However, in order to ensure that the
diffusive flux vanishes, we must require that $dg/d\tau \to 0$ as $\tau
\to 0$. Conversely, in the upstream region, we expect that $g \to 0$ as
$\tau \to \infty$ since no photons can diffuse to large distances in the
direction opposing the plasma flow. With these boundary conditions taken
into consideration, we find that the fundamental solution for the
spatial separation function $g$ has the general form
\begin{equation}
g(\lambda,\tau) \propto \cases{
e^{-\alpha(3+w)\tau^2/4} \, M\left(a,{1 \over 2},{\alpha w \tau^2
\over 2}\right) \ , & $\tau \le \tau_0$ \ , \cr
e^{-\alpha(3+w)\tau^2/4} \, U\left(a,{1 \over 2},{\alpha w \tau^2
\over 2}\right) \ , & $\tau \ge \tau_0$ \ , \cr
}
\label{eq4.6}
\end{equation}
where $M$ and $U$ denote confluent hypergeometric functions (Abramowitz
\& Stegun 1970), and we have made the definitions
\begin{equation}
a \equiv {w + 3 - 2 \lambda \over 4 w} \ ,
\ \ \ \ \
w \equiv \left(9 + 12 \, \xi^2\right)^{1/2}
\ .
\label{eq4.7}
\end{equation}
Note that in the radiation-dominated case with $\xi = 2/\sqrt{3}$, we obtain
$w=5$.

Equation~(\ref{eq4.3}) is linear, second-order, and homogeneous, and
consequently both the function $g$ and its derivative $dg/d\tau$ must be
continuous at the source location, $\tau=\tau_0$. The smooth merger of
the $M$ and $U$ functions at the source location requires that their
Wronskian,
\begin{eqnarray}
\omega(\tau) &\equiv& M\left(a,{1 \over 2},{\alpha w \tau^2 \over 2}\right)
{d \over d\tau}\,U\left(a,{1 \over 2},{\alpha w \tau^2 \over 2}\right)
\nonumber
\\
&-& \ U\left(a,{1 \over 2},{\alpha w \tau^2 \over 2}\right)
{d \over d\tau}\,M\left(a,{1 \over 2},{\alpha w \tau^2 \over 2}\right)
\ ,
\label{eq4.8}
\end{eqnarray}
must vanish at $\tau=\tau_0$. This condition can be used to solve for
the eigenvalues of the separation constant $\lambda$. By employing
equation~(13.1.22) from Abramowitz \& Stegun (1970) to evaluate the
Wronskian, we obtain the eigenvalue equation
\begin{eqnarray}
\omega(\tau)
= - {\Gamma(1/2) \, (2 \alpha w)^{1/2} \over \Gamma(a)}
\ e^{\alpha w \tau^2/2} = 0
\ .
\label{eq4.9}
\end{eqnarray}
The left-hand side vanishes when $\Gamma(a) \to \pm \infty$, which
implies that $a=-n$, where $n=0,1,2,$\ldots By combining this result
with equation~(\ref{eq4.7}), we conclude that the eigenvalues
$\lambda_n$ are given by
\begin{equation}
\lambda_n = {4nw + w + 3 \over 2} \ ,
\ \ \ \ n=0,1,2,\ldots
\label{eq4.10}
\end{equation}

When $\lambda=\lambda_n$, the spatial separation functions $g$ reduce to
a set of global eigenfunctions that satisfy the boundary conditions at
large and small values of $\tau$. In this case, we can use
equations~(13.6.9) and (13.6.27) from Abramowitz \& Stegun to show that
the confluent hypergeometric functions $M$ and $U$ appearing in
equation~(\ref{eq4.8}) are proportional to the generalized Laguerre
polynomials $L_n^{(-1/2)}$, and consequently the spatial eigenfunctions
can be written as
\begin{equation}
g_n(\tau) \equiv g(\lambda_n,\tau)
= e^{-\alpha(3+w)\tau^2/4} \, L_n^{(-1/2)}\left(\alpha w \tau^2 \over 2
\right)
\ .
\label{eq4.11}
\end{equation}
Based on equation~(7.414.3) from Gradshteyn \& Ryzhik (1980), we note
that the spatial eigenfunctions satisfy the orthogonality relation
\begin{equation}
\int_0^\infty e^{3\alpha\tau^2/2} g_n(\tau) \, g_m(\tau)
\, d\tau = \cases{
{\Gamma(n+1/2) \over n! \, \sqrt{2 \alpha w}} \ , & $n = m$ \ , \cr
0 \ , & $n \ne m$ \ . \cr
}
\label{eq4.12}
\end{equation}

\subsection{Energy Eigenfunctions}

The solution for the energy separation function $h$ depends on the
boundary conditions imposed in the energy space. As $\epsilon \to 0$, we
require that $h$ not increase faster than $\epsilon^{-3}$ since the
Green's function must possess a finite total photon number density (see
eq.~[\ref{eq3.8}]). Conversely, as $\epsilon \to \infty$, we require
that $h$ decrease more rapidly than $\epsilon^{-4}$ in order to ensure
that the Green's function contains a finite total photon energy density.
Furthermore, in order to avoid an infinite diffusive flux in the energy
space at $\chi=\chi_0$, the function $h$ must be continuous there. The
fundamental solution for the energy eigenfunction that satisfies the
various boundary and continuity conditions can be written as
\begin{equation}
h_n(\chi) \equiv h(\lambda_n,\chi) = \cases{
\chi^{\kappa-4} \, e^{-\chi/2} \, W_{\kappa,\mu}(\chi_0)
\, M_{\kappa,\mu}(\chi) \ ,
& $\chi \le \chi_0$ \ , \cr
\chi^{\kappa-4} \, e^{-\chi/2} \, M_{\kappa,\mu}(\chi_0)
\, W_{\kappa,\mu}(\chi) \ ,
& $\chi \ge \chi_0$ \ , \cr
}
\label{eq4.13}
\end{equation}
where $M_{\kappa,\mu}$ and $W_{\kappa,\mu}$ denote Whittaker functions,
and we have made the definitions
\begin{equation}
\kappa \equiv {1 \over 2} \, (\deltapar+4) \ ,
\ \ \ \ \
\mu \equiv {1 \over 2} \left[(3-\deltapar)^2 + 4 \, \deltapar \lambda_n \right]
^{1/2}
\ .
\label{eq4.14}
\end{equation}
Note that each of the eigenvalues $\lambda_n$ results in a different
value for $\mu$, and the parameter $\deltapar$ is defined in
equation~(\ref{eq4.5}). Equation~(\ref{eq4.13}) can also be written in
the more compact form
\begin{equation}
h_n(\chi) = \chi^{\kappa-4} \, e^{-\chi/2} \, M_{\kappa,\mu}(\chimin)
\, W_{\kappa,\mu}(\chimax)
\ ,
\label{eq4.15}
\end{equation}
where
\begin{equation}
\chimin \equiv \min(\chi,\chi_0) \ ,
\ \ \ \ \
\chimax \equiv \max(\chi,\chi_0)
\ .
\label{eq4.16}
\end{equation}

\subsection{Eigenfunction Expansion}

The spatial eigenfunctions $g_n(y)$ form an orthogonal set, as expected since
this is a standard Sturm-Liouville problem. The solution for the Green's
function can therefore be expressed as the infinite series
\begin{equation}
\green(\tau_0,\tau,\chi_0,\chi)
= \sum_{n=0}^\infty \ C_n \, g_n(\tau) \, h_n(\chi)
\ ,
\label{eq4.17}
\end{equation}
where the expansion coefficients $C_n$ are computed by employing the
orthogonality of the eigenfunctions, along with the derivative jump
condition
\begin{equation}
\lim_{\varepsilon \to 0} \
{\partial\green \over \partial\chi}\Bigg|_{\chi=\chi_0+\varepsilon}
- {\partial\green \over \partial\chi}\Bigg|_{\chi=\chi_0-\varepsilon}
= - \, {3 \dot N_0 \, \deltapar \, k T_e \, \delta(\tau-\tau_0) \over
\alpha \, \pi r_0^2 \, c \, \epsilon_0^4}
\ ,
\label{eq4.18}
\end{equation}
which is obtained by integrating the transport equation~(\ref{eq4.1})
with respect to $\chi$ in a very small range surrounding the injection
energy $\chi_0$. Substituting using the expansion for $\green$ yields
\begin{equation}
\lim_{\varepsilon \to 0} \
\sum_{n=0}^\infty \ C_n \, g_n(\tau) \, \left[h_n'(\chi_0+\varepsilon)
- h_n'(\chi_0-\varepsilon)\right]
= - \, {3 \dot N_0 \, \deltapar \, k T_e \, \delta(\tau-\tau_0) \over
\alpha \, \pi r_0^2 \, c \, \epsilon_0^4}
\ ,
\label{eq4.19}
\end{equation}
where primes denote differentiation with respect to $\chi$. By employing
equation~(\ref{eq4.15}) for $h_n$, we find that
\begin{equation}
\sum_{n=0}^\infty \ C_n \, g_n(\tau) \, \chi_0^{\kappa-4}
\, e^{-\chi_0/2} \mathfrak W(\chi_0)
= - \, {3 \dot N_0 \, \deltapar \, k T_e \, \delta(\tau-\tau_0) \over
\alpha \, \pi r_0^2 \, c \, \epsilon_0^4}
\ ,
\label{eq4.20}
\end{equation}
where the Wronskian, $\mathfrak W$, is defined by
\begin{equation}
\mathfrak W(\chi_0) \equiv M_{\kappa,\mu}(\chi_0) \,
W'_{\kappa,\mu}(\chi_0) - W_{\kappa,\mu}(\chi_0) \,
M'_{\kappa,\mu}(\chi_0)
\ .
\label{eq4.21}
\end{equation}

The Wronskian can be evaluated analytically using equations~(13.1.22),
(13.1.32), and (13.1.33) from Abramowitz \& Stegun (1970), which yields
\begin{equation}
\mathfrak W(\chi_0) = - \, {\Gamma(1+2\mu) \over \Gamma(\mu-\kappa+1/2)}
\ .
\label{eq4.22}
\end{equation}
Using this result to substitute for $\mathfrak W(\chi_0)$ in
equation~(\ref{eq4.20}) and reorganizing the terms, we obtain
\begin{equation}
\sum_{n=0}^\infty \ {\Gamma(1 + 2 \mu) \, C_n \, g_n(\tau) \over
\Gamma(\mu-\kappa+1/2)}
= {3 \dot N_0 \, \deltapar \, e^{\chi_0/2} \, \delta(\tau-\tau_0) \over
\alpha \, \pi r_0^2 \, c \, \chi_0^\kappa \, (k T_e)^3}
\ ,
\label{eq4.23}
\end{equation}
where $\mu$ is a function of $\lambda_n$ via equation~(\ref{eq4.14}). We
can now calculate the expansion coefficients $C_n$ by utilizing the
orthogonality of the spatial eigenfunctions $g_n$ represented by
equation~(\ref{eq4.12}). Multiplying both sides of
equation~(\ref{eq4.23}) by $e^{3\alpha\tau^2/2} g_m(\tau)$ and
integrating with respect to $\tau$ from zero to infinity yields, after
some algebra,
\begin{equation}
C_n = {3 \dot N_0 \, \deltapar \, \sqrt{2 w}
\ e^{\chi_0/2} \, e^{3\alpha\tau_0^2/2}
\over \pi r_0^2 \, c (k T_e)^3
\chi_0^\kappa \, \sqrt{\alpha}}
\, {\Gamma(\mu-\kappa+1/2) \ n! \,  g_n(\tau_0)
\over \Gamma(1+2\mu) \, \Gamma(n+1/2)}
\ .
\label{eq4.24}
\end{equation}
The final closed-form solution for the Green's function, obtained by
combining equations~(\ref{eq4.15}), (\ref{eq4.17}), and (\ref{eq4.24}),
is given by
\begin{eqnarray}
\green(\tau_0,\tau,\chi_0,\chi)
&=& {3 \dot N_0 \, \deltapar \, e^{3\alpha\tau_0^2/2} \sqrt{2 w}
\ \chi^{\kappa-4} \, e^{(\chi_0-\chi)/2} \over
\pi r_0^2 \, c (k T_e)^3 \chi_0^\kappa \sqrt{\alpha}}
\sum_{n=0}^\infty \
{\Gamma(\mu-\kappa+1/2) \, n! \over \Gamma(1+2\mu) \, \Gamma(n+1/2)}
\nonumber
\\
&\times& g_n(\tau_0) \, g_n(\tau)
\, M_{\kappa,\mu}(\chimin)
\, W_{\kappa,\mu}(\chimax)
\ ,
\label{eq4.25}
\end{eqnarray}
where the spatial eigenfunctions $g_n$ are computed using
equation~(\ref{eq4.11}) and the parameters $\kappa$ and $\mu$ are given
by equations~(\ref{eq4.14}). The Green's function can also be expressed
directly in terms of the photon energy $\epsilon$ by writing
\begin{eqnarray}
\green(\tau_0,\tau,\epsilon_0,\epsilon)
&=& {3 \dot N_0 \, \deltapar \, k T_e \, e^{3\alpha\tau_0^2/2}
\sqrt{2 w} \ \epsilon^{\kappa-4} \, e^{(\epsilon_0-\epsilon)/(2kT_e)} \over
\pi r_0^2 \, c \, \epsilon_0^\kappa \sqrt{\alpha}}
\sum_{n=0}^\infty \
{\Gamma(\mu-\kappa+1/2) \, n! \over \Gamma(1+2\mu) \, \Gamma(n+1/2)}
\nonumber
\\
&\times& g_n(\tau_0) \, g_n(\tau)
\, M_{\kappa,\mu}\left(\epsmin \over k T_e\right)
W_{\kappa,\mu}\left(\epsmax \over k T_e\right)
\ ,
\label{eq4.26}
\end{eqnarray}
where
\begin{equation}
\epsmin \equiv \min(\epsilon,\epsilon_0) \ ,
\ \ \ \ \
\epsmax \equiv \max(\epsilon,\epsilon_0)
\ .
\label{eq4.27}
\end{equation}
This exact, analytical solution for $\green$ is one of the main results
of the paper, and it provides a very efficient means for computing the
steady-state Green's function resulting from the continual injection of
monochromatic seed photons from a source at an arbitrary location inside
the accretion column. The eigenfunction expansion converges rapidly and
therefore we can generally obtain an accuracy of at least four
significant figures in our calculations of $\green$ by terminating the
series in equations~(\ref{eq4.25}) or (\ref{eq4.26}) after the first 5-10
terms.

\subsection{Asymptotic Power-Law Behavior}

Asymptotic analysis of the energy function $W_{\kappa,\mu}$ appearing in
equation~(\ref{eq4.26}) reveals that if the photon energy $\epsilon$ and
injection energy $\epsilon_0$ satisfy the conditions $\epsilon_0 < \epsilon
\ll kT_e$, then the spectrum has the power-law form
\begin{equation}
\green \propto \epsilon^{-[3-\deltapar+\sqrt{(3-\deltapar)^2
+ 4\lambda_0\deltapar}]/2} \ , \ \ \ \ \
\lambda_0 = {w + 3 \over 2}
\ ,
\label{eq4.28}
\end{equation}
where $\deltapar$ is given by equation~(\ref{eq4.5}) and $\lambda_0$ is
the leading eigenvalue computed using equation~(\ref{eq4.10}) with
$n=0$. We will demonstrate in \S~6 that bulk Comptonization dominates
over thermal Comptonization in the limit of large $\deltapar$, in which
case equation~(\ref{eq4.28}) reduces to
\begin{equation}
\green \propto \epsilon^{-\lambda_0} \ , \ \ \ \ \
\deltapar\to\infty
\ ,
\label{eq4.29}
\end{equation}
which is the same solution obtained by Becker \& Wolff (2005b) in the
case of pure bulk Comptonization. However, in contrast to the pure-bulk
model, in the scenario considered here the power-law shape only extends
up to the Wien turnover at photon energy $\epsilon \sim kT_e$.

We can gain some insight into the role of thermal Comptonization by
comparing the power-law indices in equations~(\ref{eq4.28}) and
(\ref{eq4.29}). In general, we find that
\begin{equation}
\lambda_0 \ \ge \ {3-\deltapar+\sqrt{(3-\deltapar)^2
+ 4\lambda_0\deltapar} \over 2}
\ ,
\label{eq4.30}
\end{equation}
with the equality holding in the limit of large $\deltapar$. This
implies that for small values of $\deltapar$, thermal Comptonization
causes a flattening of the spectrum in the region $\epsilon \ll kT_e$
due to the transfer of energy from high- to medium-energy photons via
electron recoil. If the flow is radiation-dominated, as expected in the
bright pulsars, then $\xi=2/\sqrt{3}$ and we obtain $w=5$ and
$\lambda_0=4$ according to equations~(\ref{eq4.7}) and (\ref{eq4.28}),
respectively. Luminous pulsars are generally dominated by bulk
Comptonization, and therefore it follows that the photon spectral index
$\alpha_{\rm X} \equiv \lambda_0-2=2$ in the region of the spectrum
below the Wien cutoff.

\section{SPECTRUM OF THE ESCAPING RADIATION}

Equations~(\ref{eq4.25}) and (\ref{eq4.26}) represent the exact solution
for the Green's function $\green$ describing the radiation spectrum {\it
inside} a pulsar accretion column resulting from the injection of $\dot
N_0$ seed photons per unit time from a monochromatic source located at
$\tau=\tau_0$ (or, equivalently, at $z=z_0$). Since the fundamental
transport equation~(\ref{eq3.1}) is linear, we can use the analytical
solutions for the Green's function to calculate the spectrum of the
radiation escaping from the accretion column for any desired source
distribution.

\subsection{Green's Function for the Escaping Radiation Spectrum}

In the escape-probability approach employed here, the associated Green's
function for the number spectrum of the photons {\it escaping through
the walls} of the cylindrical column is computed using
\begin{equation}
\greenphoton(z_0,z,\epsilon_0,\epsilon) \equiv {\pi \, r_0^2 \,
\epsilon^2 \over t_{\rm esc}(z)} \, \green(z_0,z,\epsilon_0,\epsilon)
\ ,
\label{eq5.1}
\end{equation}
where the escape timescale $t_{\rm esc}$ is evaluated as a function of
$z$ by combining equations~(\ref{eq3.4}) and (\ref{eq3.15}), which yields
\begin{equation}
t_{\rm esc}(z) = \left(\dot M \, \sigperp^2 \, r_0^2 \over 2 \pi \, m_p
\, c^3 \, \sigpar \, \alpha \, z\right)^{1/2}
\ .
\label{eq5.2}
\end{equation}
The quantity $\greenphoton \, dz \, d\epsilon$ represents the number of
photons emitted from the disk-shaped volume between positions $z$ and $z
+ dz$ per unit time with energy between $\epsilon$ and $\epsilon +
d\epsilon$.

Since the quantities $(z,z_0)$ and $(\tau,\tau_0)$ are interchangeable
via equations~(\ref{eq3.10}), we are free to work in terms of the more
convenient parameters $(\tau,\tau_0)$ without loss of generality. In
this case equations~(\ref{eq3.12}), (\ref{eq3.14}), and (\ref{eq5.2})
can be combined to reexpress the escape timescale as
\begin{equation}
t_{\rm esc}(\tau) = {r_0 \over \alpha \, \xi \, c \, \tau}
\left(\sigperp \over \sigpar\right)^{1/2}
\ .
\label{eq5.3}
\end{equation}
By using this result to substitute for $t_{\rm esc}$ in
equation~(\ref{eq5.1}), we find that the Green's function for the
escaping radiation spectrum is given in terms of the optical depth by
\begin{equation}
\greenphoton(\tau_0,\tau,\epsilon_0,\epsilon)
= \pi r_0 \, c \, \alpha \, \xi
\left(\sigpar\over\sigperp\right)^{1/2}
\tau \, \epsilon^2 \, \green(\tau_0,\tau,\epsilon_0,\epsilon)
\ ,
\label{eq5.4}
\end{equation}
where $\green$ is computed using equation~(\ref{eq4.26}). The factor of
$\tau$ on the right-hand side of equation~(\ref{eq5.4}) indicates that
the emitted radiation is strongly attenuated near the stellar surface
due to the divergence of the electron number density, which inhibits the
escape of the photons through the walls of the accretion column.
Equations~(\ref{eq4.26}) and (\ref{eq5.4}) can be combined to express
the closed-form solution for $\greenphoton$ as
\begin{eqnarray}
&\phantom{a}&
\greenphoton(\tau_0,\tau,\epsilon_0,\epsilon)
= {3 \dot N_0 \, \deltapar \, \xi \, k T_e \, \sqrt{2 \alpha w \sigpar}
\ e^{3\alpha\tau_0^2/2}
\epsilon^{\kappa-2}
\, e^{(\epsilon_0-\epsilon)/(2kT_e)} \, \tau \over
r_0 \, \epsilon_0^\kappa \, \sqrt{\sigperp}}
\nonumber
\\
&\phantom{a}&
\qquad
\times \sum_{n=0}^\infty \
{\Gamma(\mu-\kappa+1/2) \, n! \over \Gamma(1+2\mu) \, \Gamma(n+1/2)}
\ g_n(\tau_0) \, g_n(\tau)
\, M_{\kappa,\mu}\left(\epsmin \over k T_e\right)
W_{\kappa,\mu}\left(\epsmax \over k T_e\right)
\ ,
\label{eq5.5}
\end{eqnarray}
where $\epsmin$ and $\epsmax$ are defined by equations~(\ref{eq4.27}),
and $w$, $\kappa$, and $\mu$ are computed using equations~(\ref{eq4.7})
and (\ref{eq4.14}).

\begin{figure}[t]
\begin{center}
\epsfig{file=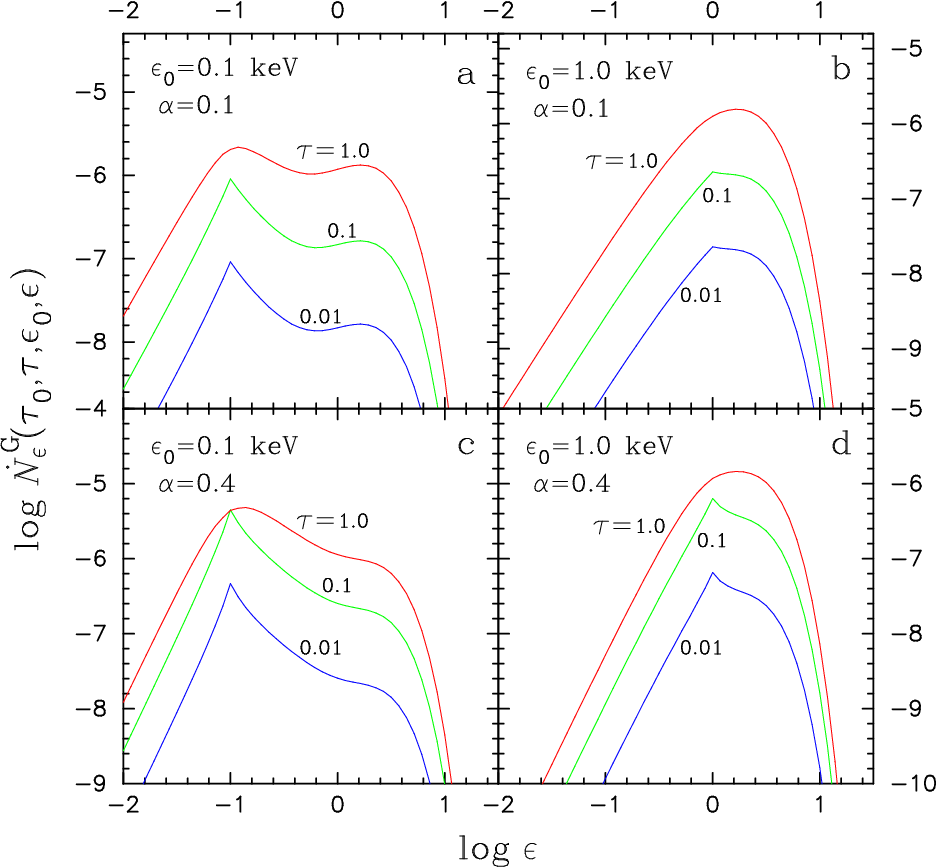,height=10.0cm}
\end{center}
\caption{Green's function $\greenphoton$ ($\rm
s^{-1}\,cm^{-1}\,keV^{-1}$) describing the photon number spectrum
escaping from an X-ray pulsar accretion column as a function of the
radiation energy $\epsilon$ (keV) and the scattering optical depth
$\tau$ above the surface, evaluated using eq.~(\ref{eq5.5}). The curves
were computed by setting $\sigpar=10^{-3}\,\sig$, $\sigperp=\sig$,
$\sigbar=0.1\,\sig$, $\dot N_0=1$, $\xi=2/\sqrt{3}$, $\tau_0=0.1$,
$r_0=10^4\,$cm, and $T_e=10^7\,$K. The values of $\tau$, $\epsilon_0$,
and $\alpha$ are indicated for each plot, and the values of $\deltapar$
are discussed in the text.}
\end{figure}

In Figure~3 we plot the Green's function for the escaping photon number
distribution, $\greenphoton$, as a function of the photon energy
$\epsilon$ and the scattering optical depth $\tau$ above the stellar
surface for two values of the parameters $\alpha$ and $\epsilon_0$ using
equation~(\ref{eq5.5}). In order to clearly illustrate the general
features of the spectrum, we set $\sigpar=10^{-3}\,\sig$,
$\sigperp=\sig$, $\sigbar=0.1\,\sig$, $\dot N_0=1$, $\xi=2/\sqrt{3}$,
$\tau_0=0.1$, $r_0=10^4\,$cm, and $T_e=10^7\,$K. We show in \S~6 that
the parameter $\deltapar$ defined by equation~(\ref{eq4.5}) determines
the relative importance of bulk and thermal Comptonization. The values
of $\deltapar$ obtained in Figure~3 are $\deltapar=0.20$ for
$\alpha=0.1$, and $\deltapar=0.79$ for $\alpha=0.4$. For the cases with
$\epsilon_0=0.1\,$keV, a distinct Wien hump is visible at $\epsilon \sim
1\,$keV due to thermal Comptonization. However, the hump is less
pronounced when $\alpha=0.4$ due to the greater strength of bulk
Comptonization, which tends to create a power-law spectral shape with
photon index $\alpha_{\rm X}=2$ (see \S~4.4). The Wien feature is almost
invisible when $\epsilon_0=1\,$keV because in this case the energy of
the injected photons is comparable to the thermal energy of the
electrons. Note that for small values of $\tau$, the escape of radiation
is inhibited because of the divergence of the electron density near the
stellar surface. Little radiation is able to escape from the column for
$\tau \gg 1$ because the photons are advectively trapped near the
stellar surface due to the high-speed inflow (see \S~6). Hence the
spatial distribution of the escaping radiation peaks around $\tau \sim
1$.

\subsection{Altitude-Dependent Spectrum for an Arbitrary Source}

Since the fundamental transport equation~(\ref{eq3.1}) is linear, the
analytical results for the Green's function obtained in \S~5.1 provide
the basis for the consideration of any source distribution. By analogy
with equation~(\ref{eq5.1}), we can compute the photon spectrum emitted
through the walls of the accretion column for an arbitrary source
distribution $Q$ in equation~(\ref{eq3.1}) by writing
\begin{equation}
\dot N_\epsilon(z,\epsilon) \equiv {\pi \, r_0^2 \,
\epsilon^2 \over t_{\rm esc}(z)} \, f(z,\epsilon)
\ ,
\label{eq5.6}
\end{equation}
where $f$ is the particular solution calculated using the integral
convolution given by equation~(\ref{eq3.6}). Substituting for $\green$
and $f$ in equation~(\ref{eq3.6}) using equations~(\ref{eq5.1}) and
(\ref{eq5.6}), respectively, we find that the particular solution for
the emitted photon spectrum can be written as
\begin{equation}
\dot N_\epsilon(z,\epsilon) = \int_0^\infty\int_0^\infty
{\greenphoton(z_0,z,\epsilon_0,\epsilon) \over \dot N_0} \ \epsilon_0^2
\, Q(z_0,\epsilon_0) \, d\epsilon_0 \, dz_0
\ .
\label{eq5.7}
\end{equation}
The spatial integration converges despite the infinite upper bound
for $z_0$ because the seed photon distribution $Q$ is localized
in the lower region of the accretion column (see \S~6.1).
Equation~(\ref{eq5.7}) facilitates the calculation of the photon
spectrum emitted by the accretion column as a function of photon energy
$\epsilon$ and altitude $z$ for any source distribution $Q$. However,
due to the large distances to the known pulsars, current observations
are unable to resolve the spatial distribution of the emission from the
column. It is therefore necessary to integrate the emitted radiation
field with respect to $z$ in order to compare the theoretical
predictions with the available spectral data, as we discuss below.

\subsection{Column-Integrated Escaping Green's Function}

By integrating over the vertical structure of the accretion column, we
can compute the total emitted radiation distribution, which corresponds
approximately to the phase-averaged spectrum of the X-ray pulsar. In the
cylindrical geometry employed here, the formal integration domain is the
region $0< z < \zmax$, where $\zmax$ is the altitude at the upper
surface of the radiating region within the accretion column, determined
by setting the inflow velocity equal to the local free-fall velocity
(see \S~6.1). In practice, however, we can replace the upper integration
bound $\zmax$ with infinity without introducing any significant error
because very few photons escape more than one or two scattering optical
depths above the stellar surface (see \S~6.5 and Becker 1998). The
replacement of the upper bound is advantageous because the resulting
integral can be performed analytically. For the case of a monochromatic
source, we therefore define the {\it column-integrated Green's function}
for the escaping photon spectrum using
\begin{equation}
\greencolumn(z_0,\epsilon_0,\epsilon)
\equiv \int_0^\infty \greenphoton(z_0,z,\epsilon_0,\epsilon)
\, dz
\ ,
\label{eq5.8}
\end{equation}
where $\greencolumn \, d\epsilon$ represents the total number of photons
escaping from the column per unit time with energy between $\epsilon$
and $\epsilon + d\epsilon$.
Substituting for
$\greenphoton$ using equation~(\ref{eq5.5}) and transforming
the variable of integration from $z$ to $\tau$
using equation~(\ref{eq3.10}) yields the alternative form
\begin{equation}
\greencolumn(\tau_0,\epsilon_0,\epsilon)
= \pi r_0^2 \, c \, \alpha^2 \, \xi^2 \, \epsilon^2
\int_0^\infty \tau^2 \green(\tau_0,\tau,\epsilon_0,\epsilon)
\, d\tau
\ ,
\label{eq5.9}
\end{equation}
where $\green$ is computed using equation~(\ref{eq4.26}) and we have
also employed equation~(\ref{eq3.12}). Despite the appearance of the
factor $\tau^2$ in the integrand in equation~(\ref{eq5.9}), the
contribution to the integral from large values of $\tau$ is actually
negligible because the spectrum declines exponentially in the upstream
region due to advection. Furthermore, the escaping spectrum is also
strongly attenuated in the downstream region ($\tau\to 0$) due to the
divergence of the electron density. Consequently most of the radiation
is emitted from the column around $\tau \sim 1$ (see Fig.~11 from
Becker~1998).

Equations~(\ref{eq4.26}) and (\ref{eq5.9}) can be combined to show that
the closed-form solution for the column-integrated Green's function is
given by
\begin{eqnarray}
&\phantom{a}&
\greencolumn(\tau_0,\epsilon_0,\epsilon)
= 3 \dot N_0 \, \deltapar \, \xi^2 \, k T_e \, \sqrt{2\alpha^3 w}
\, e^{3\alpha\tau_0^2/2} \, \epsilon_0^{-\kappa} \, \epsilon^{\kappa-2} \,
e^{(\epsilon_0-\epsilon)/(2kT_e)}
\nonumber
\\
&\phantom{a}&
\qquad\qquad
\times \sum_{n=0}^\infty \
{\Gamma(\mu-\kappa+1/2) \, n! \, X_n \over \Gamma(1+2\mu) \, \Gamma(n+1/2)}
\ \ g_n(\tau_0)
\, M_{\kappa,\mu}\left(\epsmin \over k T_e\right)
W_{\kappa,\mu}\left(\epsmax \over k T_e\right)
\ ,
\label{eq5.10}
\end{eqnarray}
where $\epsmin$, $\epsmax$, $w$, $\kappa$, and $\mu$ are defined by
equations~(\ref{eq4.7}), (\ref{eq4.14}), and (\ref{eq4.27}),
and we have made the definition
\begin{equation}
X_n \equiv \int_0^\infty \tau^2 g_n(\tau) \, d\tau
\ .
\label{eq5.11}
\end{equation}
The integral $X_n$ can be evaluated analytically by substituting for
$g_n$ using equation~(\ref{eq4.11}) and employing equation~(7.414.7) from
Gradshteyn \& Ryzhik (1980) and equation~(15.3.3) from Abramowitz \&
Stegun (1970). After some algebra, the result obtained is
\begin{equation}
X_n = {2 \, \Gamma(n+1/2) \, (3-w)^{n-1} (3-w-4nw) \over
n! \ \alpha^{3/2} (3+w)^{n+3/2}}
\ .
\qquad
\label{eq5.12}
\end{equation}

\begin{figure}[t]
\begin{center}
\epsfig{file=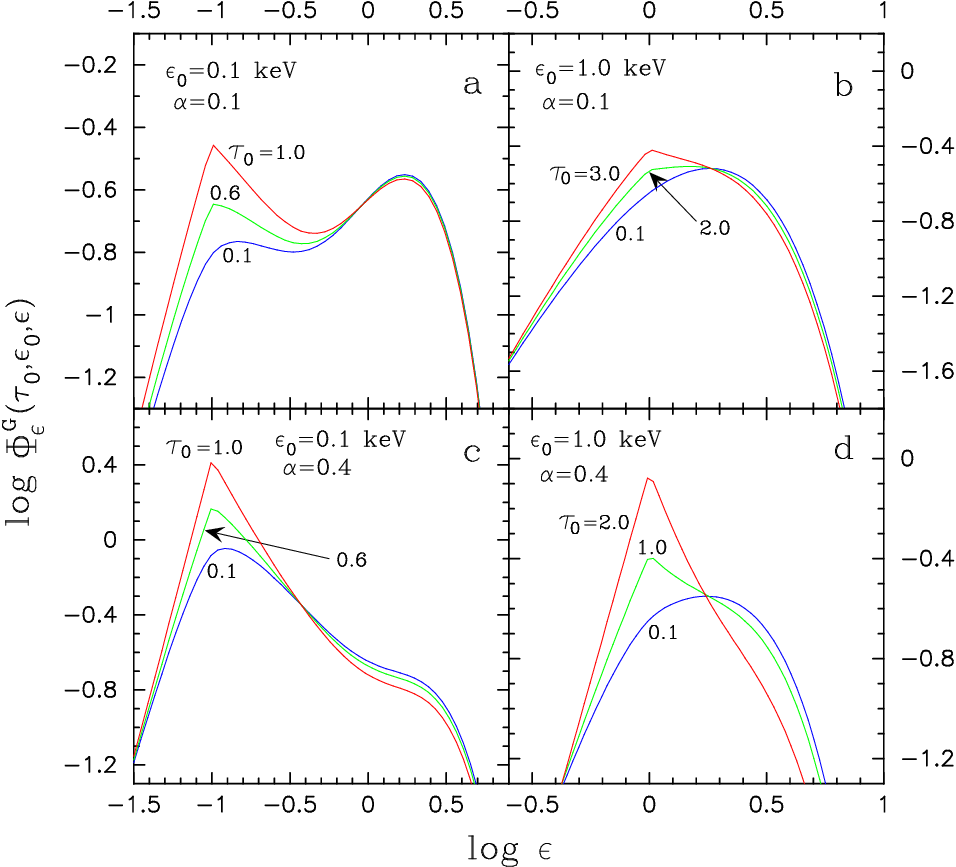,height=10.0cm}
\end{center}
\caption{Column-integrated Green's function $\greencolumn$ ($\rm
s^{-1}\,keV^{-1}$) describing the photon number spectrum escaping from
an accretion column as a function of the radiation energy $\epsilon$
(keV) and the source optical depth $\tau_0$, evaluated using
eq.~(\ref{eq5.10}). The curves were computed by setting
$\sigpar=10^{-3}\,\sig$, $\sigperp=\sig$, $\sigbar=0.1\,\sig$, $\dot
N_0=1$, $\xi=2/\sqrt{3}$, $r_0=10^4\,$cm, and $T_e=10^7\,$K. The values
of $\tau_0$, $\epsilon_0$, and $\alpha$ are indicated for each plot.}
\end{figure}

In Figure~4 we use equation~(\ref{eq5.10}) to plot the dependence of the
column-integrated Green's function $\greencolumn$ on the radiation
energy $\epsilon$ using the same parameter values employed in Figure~3,
along with several values for the source optical depth $\tau_0$, where
the monochromatic seed photons are injected. In each case, we set
$\sigpar=10^{-3}\,\sig$, $\sigperp=\sig$, $\sigbar=0.1\,\sig$, $\dot
N_0=1$, $\xi=2/\sqrt{3}$, $r_0=10^4\,$cm, and $T_e=10^7\,$K. For the
cases with $\epsilon_0=0.1\,$keV, thermal Comptonization produces a Wien
hump at $\epsilon \sim 1\,$keV, which is the same behavior noted in
Figure~3. When $\alpha=0.4$, the Wien hump is less prominent due to the
effect of bulk Comptonization. In general, smaller values of the source
optical depth $\tau_0$ result in more thermal Comptonization because the
scattering plasma is more dense. Conversely, for large values of
$\tau_0$, bulk Comptonization tends to produce a power-law spectral
shape around the photon injection energy. In each case, numerical
integration of $\greencolumn$ with respect to the radiation energy
$\epsilon$ confirms that the total number of photons escaping from the
column per unit time is equal to unity, which is correct since we have
set $\dot N_0=1$.

\subsection{Column-Integrated Spectrum for an Arbitrary Source}

The vertically-integrated photon spectrum emitted through the walls of
the accretion column due to an arbitrary source $Q$ is given by (cf.
eq.~[\ref{eq5.8}])
\begin{equation}
\Phi_\epsilon(\epsilon)
\equiv \int_0^\infty \dot N_\epsilon(z,\epsilon)
\, dz
\ ,
\label{eq5.13}
\end{equation}
where $\dot N_\epsilon$ is the altitude-dependent particular solution
for the escaping photon spectrum computed using equation~(\ref{eq5.7}).
By substituting for $\dot N_\epsilon$ using equation~(\ref{eq5.7}),
interchanging the order of integration, and applying
equation~(\ref{eq5.8}), we find that the particular solution for the
column-integrated spectrum can be written as
\begin{equation}
\Phi_\epsilon(\epsilon) = \int_0^\infty\int_0^\infty
{\greencolumn(z_0,\epsilon_0,\epsilon) \over \dot N_0} \ \epsilon_0^2
\, Q(z_0,\epsilon_0) \, d\epsilon_0 \, dz_0
\ ,
\label{eq5.14}
\end{equation}
where $\greencolumn$ is evaluated using equation~(\ref{eq5.10}), with
$\tau_0$ computed in terms of $z_0$ using equation~(\ref{eq3.14}).
Equation~(\ref{eq5.14}) facilitates the calculation of the total photon
spectrum emitted by the entire column as a function of the photon energy
$\epsilon$, for any source $Q$. In \S~7 we examine the nature of the
source term for the various emission processes important in X-ray pulsar
accretion columns, and we derive associated results for the
column-integrated escaping photon spectrum, $\Phi_\epsilon$.
Astrophysical applications of these results and comparisons with the
X-ray data for specific pulsars are presented in \S~8.

\section{MODEL PARAMETERS AND CONSTRAINTS}

Our model includes a number of free parameters, such as the column
radius, $r_0$, the temperature of the gas in the thermal mound,
$\Tmound$, the temperature of the electrons in the optically thin region
above the mound, $T_e$, the magnetic field strength, $B$, and the
accretion rate, $\dot M$. There are also several additional theory
parameters that are related to the above mentioned physical parameters,
including $\alpha$, $\deltapar$, $\xi$, and the scattering cross
sections $\sigperp$, $\sigpar$, and $\sigbar$. In this section we
investigate the relationships between these various quantities, and we
also introduce several new constraints that are used to reduce the
number of free parameters. Based on these results, we show that for
given values of the stellar mass $M_*$ and radius $R_*$, the only
quantities that need to be varied when fitting the spectral data for a
particular source are the six parameters $T_e$, $\dot M$, $r_0$, $B$,
$\deltapar$, and $\xi$.

\subsection{Dynamical Constraints}

At large distances from the star, where radiation pressure effects
become negligible, we require that the approximate velocity given by
equation~(\ref{eq3.15}) equal the local free-fall velocity. We use this
condition to calculate the altitude, $\zmax$, at the upper surface of
the radiating region within the accretion column by writing
\begin{equation}
\left(2 \, G M_* \over R_* + \zmax\right)^{1/2}
= c \, \alpha \, \taumax
\ ,
\label{eq6.1.1}
\end{equation}
where the left-hand side represents the free-fall velocity from infinity
to the top of the radiative zone. The corresponding optical depth,
$\taumax$, is related to $\zmax$ via (see eq.~[\ref{eq3.14}])
\begin{equation}
\taumax = \left(\sigpar \over \sigperp\right)^{1/4}
\left(2 \, \zmax \over \alpha \, \xi \, r_0\right)^{1/2}
\ .
\label{eq6.1.2}
\end{equation}
By using equation~(\ref{eq6.1.2}) to substitute for $\taumax$ in
equation~(\ref{eq6.1.1}), we are able to derive a quadratic equation for
$\zmax$ with solution
\begin{equation}
\zmax = {R_* \over 2} \, \left[\left(1 + C_1\right)^{1/2} - 1 \right]
\ ,
\label{eq6.1.3}
\end{equation}
where
\begin{equation}
C_1 \equiv {4 \, G M_* \, r_0 \, \xi
\over \alpha \, c^2 R_*^2} 
\left(\sigperp \over \sigpar\right)^{1/2}
\ .
\label{eq6.1.4}
\end{equation}
We will use this result to set the upper limit for the spatial integrations
performed in \S~7 when we calculate the emergent spectra resulting from
the Comptonization of bremsstrahlung and cyclotron seed photons.

\subsection{Scattering Cross Sections}

Several electron scattering cross sections are incorporated into the
model, based on the direction of propagation of the photon. The cross
sections for photons propagating parallel and perpendicular to the
magnetic field direction are denoted by $\sigpar$ and $\sigperp$,
respectively, and the angle-averaged cross section is $\sigbar$.
Following Wang \& Frank (1981), we set the perpendicular cross section
equal to the Thomson value, so that (see eq.~[\ref{eq2.7}])
\begin{equation}
\sigperp = \sig
\ .
\label{eq6.2.1}
\end{equation}
For given values of the parameters $\dot M$, $r_0$, and $\xi$, we can
compute the parallel cross section $\sigpar$ by using equation~(\ref{eq3.12})
to write
\begin{equation}
\sigpar = \left(\pi r_0 \, m_p \, c \over \dot M \, \xi \right)^2
{1 \over \sigperp}
\ .
\label{eq6.2.2}
\end{equation}
Once values are specified for the electron temperature $T_e$ and the
parameters $\alpha$ and $\deltapar$, we can compute the angle-averaged
cross section $\sigbar$ by using equation~(\ref{eq4.5}), which yields
\begin{equation}
\sigbar = {\alpha \over 3} \, {m_e c^2 \over k T_e} \,
{\sigpar \over \deltapar}
\ ,
\label{eq6.2.3}
\end{equation}
where $\sigpar$ is evaluated using equation~(\ref{eq6.2.2}). We shall
use equations~(\ref{eq6.2.1}), (\ref{eq6.2.2}), and (\ref{eq6.2.3}) to
set the values of the three scattering cross sections appearing in our
model. In general, we expect that $\sigpar \ll \sigbar \ll \sigperp$
(e.g., Canuto et al. 1971), and this result will be verified when
specific numerical models are developed. We will also confirm that the
values obtained for $\sigpar$ are reasonably consistent with the
dependence on the mean photon energy expressed by
equation~(\ref{eq2.6}).

\subsection{Thermal Mound Properties}

The blackbody surface of the thermal mound represents the effective
photosphere for photon creation and destruction in the accretion column,
which occurs mainly via free-free emission and absorption. The altitude
at the top of the mound can therefore be estimated by setting the free-free
optical thickness across the column equal to unity,
\begin{equation}
\tau^{\rm ff} \equiv r_0 \, \alpha_{\rm R}^{\rm ff} = 1
\ ,
\label{eq6.3.1}
\end{equation}
where $\alpha_{\rm R}^{\rm ff}$ denotes the Rosseland mean of the
free-free absorption coefficient. In principle, we should use an
expression for $\alpha_{\rm R}^{\rm ff}$ that accounts for the effect of
the strong magnetic field (e.g., Lauer et al. 1983). However, the
modifications introduced by the presence of the field are relatively
minor for photons with energies $\epsilon \ll kT_e$, which are the most
strongly absorbed. We can therefore estimate the value of $\alpha_{\rm
R}^{\rm ff}$ at the mound surface in the case of pure, fully-ionized
hydrogen by using equation~(5.20) from Rybicki \& Lightman (1979) to
write
\begin{equation}
\alpha_{\rm R}^{\rm ff} = 6.10 \times 10^{22} \ \Tmound^{-7/2}
\ \rhomound^2
\label{eq6.3.2}
\end{equation}
in cgs units, where $\Tmound$ and $\rhomound$ represent the temperature
and density at the top of the mound, respectively, and we have set the
Gaunt factor equal to unity.

By combining equations~(\ref{eq6.3.1}) and (\ref{eq6.3.2}), we find that
the density at the thermal mound surface is given by
\begin{equation}
\rhomound = 4.05 \times 10^{-12} \ \Tmound^{7/4} \ r_0^{-1/2}
\ .
\label{eq6.3.3}
\end{equation}
The inflow speed at the mound surface, $\vmound \equiv -v(\zmound)$,
is related to $\rhomound$ via the continuity equation~(\ref{eq3.5}),
which yields
\begin{equation}
\vmound = 7.86 \times 10^{10} \ \dot M \, r_0^{-3/2} \, \Tmound^{-7/4}
\ .
\label{eq6.3.4}
\end{equation}
We can now compute the optical depth at the top of the thermal mound
by combining equations~(\ref{eq3.15}), (\ref{eq3.18}), and (\ref{eq6.3.4}),
which yields
\begin{equation}
\taumound = 2.64 \times 10^{28} \, {\dot M \, R_* \over
M_* \, r_0^{3/2} \, \Tmound^{7/4} \, \xi}
\ .
\label{eq6.3.5}
\end{equation}
The corresponding altitude, $\zmound$, is obtained by combining
equations~(\ref{eq3.14}), (\ref{eq3.18}), and (\ref{eq6.3.5}), which
yields in cgs units
\begin{equation}
\zmound = 5.44 \times 10^{15} \, {\dot M \, R_* \over M_* \, r_0 \,
\Tmound^{7/2} \, \xi \, \sigpar}
\ .
\label{eq6.3.6}
\end{equation}
We will use $\zmound$ as the lower limit for the spatial integrations
performed in \S~7 when we calculate the emergent spectra resulting from
the Comptonization of bremsstrahlung and cyclotron seed photons.

The temperature of the gas in the thermal mound, $\Tmound$, is computed
separately from the temperature of the electrons, $T_e$, in the
optically thin region above the mound, although the two temperatures are
expected to be comparable. We calculate $\Tmound$ by setting the
downward flux of kinetic energy at the mound surface equal to the upward
flux of blackbody radiation, i.e.,
\begin{equation}
\sigma_{_{\rm SB}} \, \Tmound^4 = {1 \over 2} \, J \, \vmound^2
\ ,
\label{eq6.3.7}
\end{equation}
where
\begin{equation}
J \equiv \rhomound \, \vmound = {\dot M \over \pi r_0^2}
\label{eq6.3.7b}
\end{equation}
denotes the (constant) mass accretion flux and $\sigma_{_{\rm SB}}$ is
the Stephan-Boltzmann constant. The right-hand side of
equation~(\ref{eq6.3.7}) represents the flux of kinetic energy that must
be thermalized and radiated by the mound before the gas can merge onto
the star. By combining equations~(\ref{eq3.5}), (\ref{eq6.3.4}), and
(\ref{eq6.3.7}), we find that the mound temperature is given in cgs
units by
\begin{equation}
\Tmound = 2.32 \times 10^3 \, \dot M^{2/5} \, r_0^{-2/3}
\ .
\label{eq6.3.8}
\end{equation}
Once values for the fundamental free parameters $M_*$, $R_*$, $\dot M$,
$r_0$, and $\xi$ have been selected, the mound temperature $\Tmound$ is
evaluated using equation~(\ref{eq6.3.8}), after which the velocity
$\vmound$, optical depth $\taumound$, and altitude $\zmound$ at the
mound surface are calculated using equations~(\ref{eq6.3.4}),
(\ref{eq6.3.5}), and (\ref{eq6.3.6}), respectively.

\subsection{Bulk vs. Thermal Comptonization}

The dimensionless parameter $\deltapar$ defined by
equation~(\ref{eq4.5}) plays a central role in determining the overall
shape of the reprocessed radiation spectrum resulting from the bulk and
thermal Comptonization of the seed photons. We demonstrate here that
$\deltapar$ is essentially the ratio of the ``$y$-parameters'' for these
two processes. In the one-dimensional flow configuration under
consideration here, the mean energization rate for a photon with energy
$\epsilon$ due to the bulk compression of the decelerating electrons is
given by (e.g., Becker 1992)
\begin{equation}
\left\langle{d\epsilon \over dt}\right\rangle
\bigg|_{\rm bulk} = - \, {\epsilon \over 3}
\, {dv \over dz}
= {\alpha \, c \, n_e \sigpar \, \epsilon \over 3} 
\ ,
\label{eq6.4.1}
\end{equation}
where the final result is obtained by utilizing equation~(\ref{eq3.10}),
along with the approximate velocity profile given by
equation~(\ref{eq3.13}). The associated $y$-parameter, describing the
average fractional energy change experienced by a photon before it
escapes through the column walls, is defined by
\begin{equation}
y_{\rm bulk} \equiv {t_{\rm esc} \over \epsilon}
\, \left\langle{d\epsilon \over dt}\right\rangle\bigg|_{\rm bulk}
= {\alpha \, c \, n_e \sigpar \, t_{\rm esc} \over 3} 
\ ,
\label{eq6.4.2}
\end{equation}
where $t_{\rm esc}$ is the mean escape timescale, expressed as a function
of $z$ by equation~(\ref{eq5.2}).

Next we recall that the mean stochastic energization rate for the
thermal Comptonization of photons is given by (Rybicki \& Lightman 1979)
\begin{equation}
\left\langle{d\epsilon \over dt}\right\rangle \Bigg|_{\rm therm}
= n_e \, \sigbar \, c \ \epsilon \ {4 k T_e \over m_e c^2}
\ ,
\label{eq6.4.3}
\end{equation}
with associated $y$-parameter defined by
\begin{equation}
y_{\rm therm} \equiv {t_{\rm esc} \over \epsilon}
\, \left\langle{d\epsilon \over dt}\right\rangle\bigg|_{\rm therm}
= n_e \, \sigbar \, c \ t_{\rm esc} \ {4 k T_e \over m_e c^2}
\ .
\label{eq6.4.4}
\end{equation}
We note that $n_e \propto v^{-1} \propto z^{-1/2}$ and $t_{\rm esc}
\propto z^{-1/2}$ according to equations~(\ref{eq3.15}) and
(\ref{eq5.2}), respectively. Hence, $y_{\rm bulk}$ and $y_{\rm therm}$
each vary in proportion to $z^{-1}$, and therefore both quantities
diverge at the base of the column. It is therefore difficult to assign
global values for the two parameters that characterize the entire
accretion column.

We can gain insight into the relative importance of bulk and thermal
Comptonization by relating the ratio of the respective $y$-parameters to
the value of $\deltapar$ by combining equations~(\ref{eq4.5}),
(\ref{eq6.4.2}), and (\ref{eq6.4.3}), which yields
\begin{equation}
{\deltapar \over 4} = {y_{\rm bulk} \over y_{\rm therm}}
\ .
\label{eq6.4.5}
\end{equation}
According to this relation, bulk Comptonization dominates the photon
energization if $\deltapar \gg 1$, and thermal Comptonization dominates
if $\deltapar \ll 1$. In the intermediate regime, with $\deltapar\sim
1$, both processes contribute about equally to the energy transfer
between the electrons and the photons. However, it is important to
realize that even for large values of $\deltapar$, the thermal process
may still have a profound effect on the overall {\it shape} of the
spectrum by transferring energy from high to low frequency photons,
which produces a quasi-exponential cutoff at high energies and a
concomitant flattening of the spectrum at lower energies.

By combining equations~(\ref{eq6.4.1}) and (\ref{eq6.4.3}) with
equation~(7.36) from Rybicki \& Lightman (1979), we find that the total
(mean) photon energization rate for the thermal+bulk Comptonization model
considered here is given by
\begin{equation}
\left\langle{d\epsilon \over dt}\right\rangle \Bigg|_{\rm total}
= n_e \, \sigbar \, c \, \epsilon \left({4 k T_e \over m_e c^2}
- {\epsilon \over m_e c^2} + {\alpha \, \sigpar \over 3
\, \sigbar} \right)
\ ,
\label{eq6.4.6}
\end{equation}
where the terms on the right-hand side describe the effects of stochastic
energization, electron recoil losses, and bulk Comptonization, respectively.
The total energization rate vanishes at the critical energy
\begin{equation}
\epsilon_{\rm crit} = 4 k T_e + {\alpha \, \sigpar \over 3 \, \sigbar}
\ m_e \, c^2
\ .
\label{eq6.4.7}
\end{equation}
In the low-temperature limit, we find that bulk Comptonization balances
recoil losses at the energy
\begin{equation}
\epsilon_{\rm bulk} = {\alpha \, \sigpar \over 3 \, \sigbar}
\ m_e \, c^2 \ ,
\ \ \ \ T_e \to 0
\ ,
\label{eq6.4.8}
\end{equation}
which is the maximum photon energy achieved in the thermal+bulk model as
$T_e \to 0$. Note that in the pure bulk Comptonization model of Becker
\& Wolff (2005a,b), recoil losses are not included, and therefore the
power-law shape extends to infinite photon energy. This makes it
necessary to restrict the index of the photon spectrum to $\alpha_{\rm
X} > 2$ in order to avoid an infinite photon energy density in the
Becker \& Wolff model. However, no such restriction exists in the
thermal+bulk model considered here because recoil losses always
attenuate the spectrum at high photon energies.

\subsection{Photon Advection, Diffusion, and Escape}

The seed photons injected into the accretion column experience a variety
of physical effects as they propagate within the plasma, eventually
escaping through the column walls. In this section we will show that in
a radiation-dominated accretion column, the timescale for the escape of
radiation through the column walls is comparable to the dynamical
timescale for the accretion of the gas onto the star. In particular, we
will demonstrate that the dimensionless parameter $\xi$ defined in
equation~(\ref{eq3.12}) is proportional to the ratio of these two
timescales.

We define the dynamical timescale, $t_{\rm shock}$, for the accretion of
material from the sonic point down to the surface of the star by writing
\begin{equation}
t_{\rm shock} \equiv {\zst \over v(\zst)}
\ ,
\label{eq6.5.1}
\end{equation}
where $\zst$ is the distance between the sonic point and the stellar
surface given by equation~(\ref{eq3.17}). We can form the ratio of the
dynamical timescale $t_{\rm shock}$ to the mean escape timescale
$t_{\rm esc}$ for photons to diffuse through the walls of the accretion
column by employing equations~(\ref{eq3.4}) and (\ref{eq6.5.1}),
which yields
\begin{equation}
{t_{\rm shock} \over t_{\rm esc}}
= {\pi \, m_p \, c \, \zst \over \dot M \, \sigperp}
\ .
\label{eq6.5.2}
\end{equation}
We can substitute for $\dot M$ and $\zst$ using equations~(\ref{eq3.12})
and (\ref{eq3.17}), respectively, to obtain the alternative result
\begin{equation}
{t_{\rm shock} \over t_{\rm esc}}
= {\ln(7/3) \over 2 \sqrt{3}} \ \xi \sim 0.24 \ \xi
\ .
\label{eq6.5.3}
\end{equation}
Becker (1998) found that $\xi = 2/\sqrt{3}$ in a radiation-dominated
pulsar accretion flow, and therefore we conclude that
\begin{equation}
t_{\rm shock} \sim 0.28 \ t_{\rm esc}
\ .
\label{eq6.5.4}
\end{equation}
This result is expected, since the timescale for the photons to escape
from the column must be comparable to the dynamical timescale because the
radiation must be able to remove the kinetic energy from the flow. This
is a general property of radiative accretion flows (see, e.g., Imamura
et al. (1987).

We can gain additional insight into the processes affecting the
transport of photons through the accretion column by analyzing the
relative importance of the upward diffusion of photons along the column
axis, compared with the downward advection of photons caused by the
``trapping'' of radiation in the rapidly falling gas. The parameter
$\xi$ defined in equation~(\ref{eq3.12}) also provides a direct
measurement of the influence of these two competing effects. The mean
diffusion velocity for photons propagating parallel to the axis of the
accretion column, $w_{||}$, can be approximated by writing
\begin{equation}
w_{||} = {c \over \tau}
\ .
\label{eq6.5.5}
\end{equation}
Photon ``trapping'' occurs when the downward advective flux dominates
over the upward diffusive flux, so that the photons are essentially
confined in the lower region of the flow (Becker \& Begelman 1986).
In the model considered here, we find that the radiation is trapped if
$|v| > w_{||}$, which can be rewritten using equations~(\ref{eq3.13})
and (\ref{eq6.5.5}) as the condition $\tau > \tautrap$, where
\begin{equation}
\tautrap = {1 \over \sqrt{\alpha}}
\ .
\label{eq6.5.6}
\end{equation}

In the lower region of the column, where $\tau < \tautrap$,
diffusion is able to transport radiation effectively in the vertical
direction. We can calculate the associated ``trapping altitude'' for
photons diffusing vertically in the column by combining
equations~(\ref{eq3.14}) and (\ref{eq6.5.6}), which yields
\begin{equation}
\ztrap = {\pi r_0^2 \, c \, m_p \over 2 \dot M \sigpar}
= {r_0 \, \xi \over 2} \left(\sigperp \over \sigpar \right)^{1/2}
\ ,
\label{eq6.5.7}
\end{equation}
where the final result is obtained by substituting for $\dot M$ using
equation~(\ref{eq3.12}). It is interesting to compare the value of
$\ztrap$ with the distance from the stellar surface to the sonic
point, $\zst$, given by equation~(\ref{eq3.17}). We find that the ratio
of these two quantities is given by
\begin{equation}
{\ztrap \over \zst} = {\xi \, \sqrt{3} \over \ln(7/3)}
\ \sim \ 2.0 \ \xi
\ .
\label{eq6.5.8}
\end{equation}
Since $\xi \sim 1$ in a radiation-dominated pulsar accretion flow,
equation~(\ref{eq6.5.8}) indicates that $\zst$ and $\ztrap$ are
comparable, and therefore the photons are not able to penetrate very far
into the region above the sonic point. Hence most of the photons escape
through the walls of the column below the sonic point, with a mean
residence time equal to the dynamical time for the gas to settle onto
the stellar surface (see eq.~[\ref{eq6.5.4}]).

\subsection{Radius of the Accretion Column}

In our model, the radius $r_0$ of the accretion column (which is also
the radius of the ``hot spot'' on the stellar surface) is treated as a
free parameter. However, it is possible to constrain the value of this
radius based on dynamical considerations related to the
magnetically-channeled accretion of the gas onto the neutron star. For
example, by taking into account the effects of both the stellar rotation
and the strong magnetic field, Lamb et al. (1973) find that the radius
of the hot spot satisfies the constraint (see also Harding et al. 1984)
\begin{equation}
r_0 \lapprox R_* \left(R_* \over r_{\rm A}\right)^{1/2}
\ ,
\label{eq6.6.1}
\end{equation}
where $R_*$ is the stellar radius and $r_{\rm A}$ is the Alfv\'en
radius, given in cgs units by
\begin{equation}
r_{\rm A} = 2.6 \times 10^8
\left(B \over 10^{12}\,{\rm G}\right)^{4/7}
\left(R_* \over 10\,{\rm km}\right)^{10/7}
\left(M_* \over \msun\right)^{1/7}
\left(\xlum \over 10^{37}\,{\rm ergs \ s}^{-1}\right)^{-2/7}
\ .
\label{eq6.6.2}
\end{equation}
By combining equations~(\ref{eq2.8b}), (\ref{eq6.6.1}), and (\ref{eq6.6.2}),
we find that column radius $r_0$ is bounded by the upper limit
\begin{equation}
r_0 \lapprox 6.5 \times 10^4 \left(B \over 10^{12}\,{\rm G}\right)^{-2/7}
\left(R_* \over 10\,{\rm km}\right)^{9/14}
\left(M_* \over \msun\right)^{1/14}
\left(\dot M \over 10^{17}\,{\rm g \ s}^{-1}\right)^{1/7}
\ ,
\label{eq6.6.3}
\end{equation}
which represents an interesting constraint for our model. We will
compare our results for $r_0$ with this upper bound when specific
numerical models are developed in \S~8.

\subsection{Comparison of Exact and Approximate Velocity Profiles}

Following Lyubarskii \& Sunyaev (1982), we have adopted the approximate
velocity profile given by equation~(\ref{eq3.15}), which faciliates the
separation of the transport equation. The approximate velocity profile
was compared with the exact solution (eq.~[\ref{eq3.16}]) in Figure~2,
from which we conclude that the two functions agree fairly well in the
radiative portion of the accretion column. We can further explore the
validity of equation~(\ref{eq3.15}) by comparing the results obtained
for the emergent spectra using the exact and approximate velocity
profiles. This can be accomplished by contrasting the spectra computed
using the thermal+bulk Comptonization model developed here (based on the
approximate velocity distribution) with tho corresponding results
obtained using the ``pure'' bulk Comptonization model discussed by
Becker \& Wolff (2005b), which is based on the exact velocity profile.

In Figure~5 we display the results obtained for the column-integrated
Green's function, $\greencolumn$, calculated using either the
thermal+bulk Comptonization model (eq.~[\ref{eq5.10}]) or the pure bulk
Comptonization model (eq.~[81] from Becker \& Wolff 2005b). The
parameters values employed are the same ones adopted in Figure~3, with
$\epsilon_0=0.01\,$keV, $\alpha=0.4$, $\sigpar=10^{-3}\,\sig$,
$\sigperp=\sig$, $\sigbar=0.1\,\sig$, $\dot N_0=1$, $\xi=2/\sqrt{3}$,
$\tau_0=0.1$, and $r_0=10^4\,$cm. The electron temperature $T_e$ is set
equal to $10^7\,$K, $10^{6.3}\,$K, $10^{5.3}\,$K, or 0, and the
corresponding values for $\deltapar$ are 0.79, 3.95, 39.54, and
$\infty$, respectively (see eq.~[\ref{eq4.5}]). Note that a Wien hump
forms at energy $\epsilon \sim \epsilon_{\rm crit}$
(eq.~[\ref{eq6.4.7}]) for large values of $T_e$ due to photon energy
redistribution, and as $T_e \to 0$ the spectrum approaches the power-law
shape characteristic of bulk Comptonization. The high-energy cutoffs
displayed by the spectra in Figure~5 are the result of electron recoil
only; in actual X-ray pulsar spectra, the quasi-exponential cutoffs are
produced by a combination of electron recoil and cyclotron absorption,
as discussed in \S~8.

The Becker \& Wolff (2005b) spectrum is based on the {\it exact}
velocity solution (eq.~[\ref{eq3.16}]), and therefore it provides an
interesting test for the model considered here, which is based on the
approximate velocity profile (eq.~[\ref{eq3.15}]). In the limit $T_e \to
0$, bulk Comptonization dominates the spectral formation process, and
the column-integrated Green's function computed using
equation~(\ref{eq5.10}) agrees perfectly with the Becker \& Wolff result
up to photon energy $\epsilon=\epsilon_{\rm bulk}$, defined as the
energy at which bulk Comptonization is balanced by losses due to
electron recoil (see eq.~[\ref{eq6.4.8}]). Since recoil losses were not
included in the Becker \& Wolff (2005b) model, the resulting spectrum
extends to infinite photon energy. Similar agreement between the two
models in the low-temerature limit has been confirmed using several
other sets of parameters. These comparisons confirm that the approximate
velocity profile utilized here provides a good representation of the
effect of bulk Comptonization in the pulsar accretion column.

\begin{figure}[t]
\begin{center}
\epsfig{file=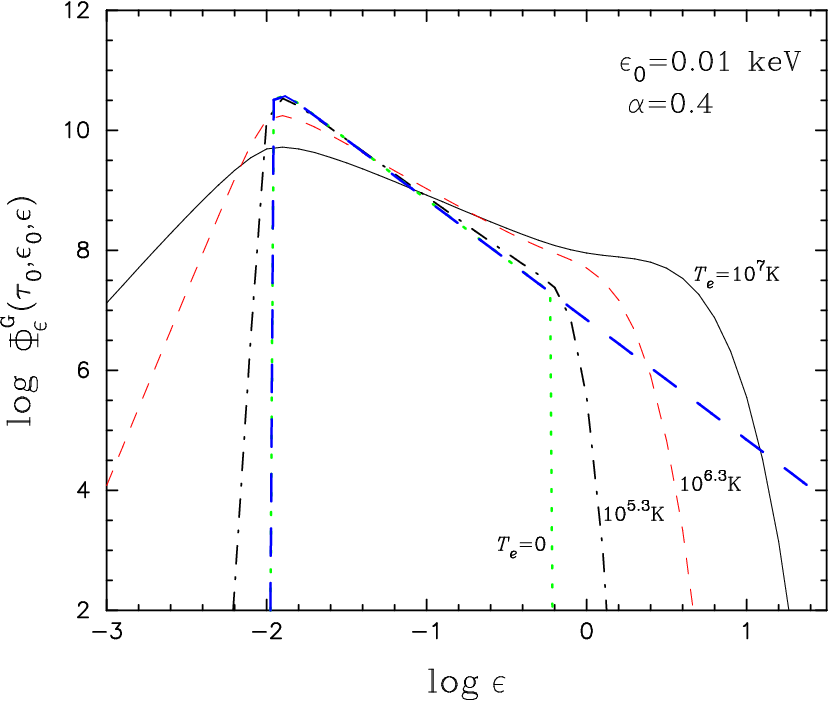,height=10.0cm}
\end{center}
\caption{Column-integrated Green's function $\greencolumn$ ($\rm
s^{-1}\,keV^{-1}$) describing the photon number spectrum escaping from
an accretion column as a function of the radiation energy $\epsilon$
(keV) and the electron temperature $T_e$, evaluated using
eq.~(\ref{eq5.10}). The results were computed by setting
$\epsilon_0=0.01\,$keV, $\alpha=0.4$, $\sigpar=10^{-3}\,\sig$,
$\sigperp=\sig$, $\sigbar=0.1\,\sig$, $\dot N_0=1$, $\xi=2/\sqrt{3}$,
$\tau_0=0.1$, and $r_0=10^4\,$cm. The corresponding value of $T_e$ is
indicated for each curve, and the heavy dashed line denotes the spectrum
obtained using the exact velocity profile; see the discussion in the
text.}
\end{figure}

\section{PHOTON SOURCES AND ASSOCIATED SPECTRA}

The high-energy radiation spectrum emerging from an X-ray pulsar
accretion column is produced via the bulk and thermal Comptonization of
seed photons injected into the plasma by a variety of emission
mechanisms. Based on the linearity of the transport
equation~(\ref{eq3.1}), we can compute the radiation spectrum emerging
from the accretion column for any source term $Q$ by employing the
integral convolutions given by equations~(\ref{eq5.7}) and
(\ref{eq5.14}) for the altitude-dependent and column-integrated spectra
$\dot N_\epsilon(z,\epsilon)$ and $\Phi_\epsilon(\epsilon)$,
respectively. The calculation of the two spectra is straightforward due
to the availability of the analytical solutions for the
altitude-dependent Green's function $\greenphoton$ (eq.~[\ref{eq5.5}])
and for the column-integrated Green's function $\greencolumn$
(eq.~[\ref{eq5.10}]). In the context of accretion-powered X-ray pulsars,
the primary sources of seed photons are bremsstrahlung, cyclotron, and
blackbody radiation (Arons et al. 1987). The first two mechanisms inject
photons throughout the column, although the emission tends to be
concentrated towards the base of the column because of the strong
density dependence of these processes. Conversely, blackbody radiation
is injected primarily at the surface of the thermal mound (i.e., the
``photosphere''), where the gas achieves thermodynamic equilibrium. The
energy dependences of the three emission mechanisms are also quite
different, with the bremsstrahlung and blackbody processes producing
broadband continuum radiation, and the cyclotron producing nearly
monochromatic radiation.

The nature of the source term $Q$ appearing in the transport
equation~(\ref{eq3.1}) is examined in detail below for each of the
emission mechanisms of interest here. In general, $Q$ can be related to
the photon emissivity, $\dot n_\epsilon$, by writing
\begin{equation}
\epsilon^2 \, Q(z,\epsilon) \, d\epsilon \, dz
= \pi r_0^2 \, \dot n_\epsilon \, d\epsilon \, dz
\ ,
\label{eq7.1}
\end{equation}
where $\dot n_\epsilon \, d\epsilon$ expresses the number of photons
created per unit time per unit volume in the energy range between
$\epsilon$ and $\epsilon+d\epsilon$. We will use equation~(\ref{eq7.1})
to determine $Q$ by specifying the emissivity $\dot n_\epsilon$ for
bremsstrahlung, cyclotron, and blackbody emission. In our approach, we
treat the cyclotron and bremsstrahlung sources as separate emission
mechanisms, although they can also be viewed as components of a single,
generalized magneto-bremsstrahlung process, as discussed by Riffert et
al. (1999).

\subsection{Cyclotron Radiation}

In an accretion-powered X-ray pulsar, cyclotron photons are produced
mainly as a result of the collisional excitation of electrons to the $n=1$
Landau level, followed by radiative decay to the ground state ($n=0$).
The electrons are excited via collisions with protons, and
therefore the production of the cyclotron radiation is a two-body
process. Estimates suggest that in the luminous sources, collisional
deexcitation is completely negligible compared with radiative
deexcitation, and therefore the production of the cyclotron radiation
acts as a cooling mechanism for the gas (e.g., Nagel 1980). The
subsequent Comptonization of these photons results in further cooling.
The cyclotron line will be broadened by thermal effects, as well as the
possible spatial variation of the magnetic field, but these effects are
likely to be masked by the strong diffusion in energy space caused by
thermal Comptonization. We can therefore approximate the distribution of
the cyclotron seed photons using a monochromatic source term. Based on
equations~(7) and (11) from Arons et al. (1987), we find that in the
case of pure, fully-ionized hydrogen, the rate of production of
cyclotron photons per unit volume per unit energy is given in cgs units
by
\begin{equation}
\dot n_\epsilon^{\rm cyc} = 2.10 \times 10^{36} \, \rho^2 \,
B_{12}^{-3/2} \, H\left(\epsilon_c \over kT_e\right) \, e^{-\epsilon_c/kT_e}
\, \delta(\epsilon-\epsilon_c)
\ ,
\label{eq7.2}
\end{equation}
where the cyclotron energy $\epsilon_c$ is defined by equation~(\ref{eq2.1}),
and
\begin{equation}
H\left(\epsilon_c \over kT_e\right) \equiv
\cases{
0.41 \ , & $\epsilon_c/kT_e > 7.5$ \ , \cr
0.15 \, \sqrt{\epsilon_c/kT_e} \ , & $\epsilon_c/kT_e < 7.5$ \ . \cr
}
\label{eq7.3}
\end{equation}
The cyclotron line will be broadened by thermal effects, as well as the
possible spatial variation of the magnetic field, but these effects are
likely to be masked by the strong diffusion in energy space caused by
thermal Comptonization. We can therefore approximate the cyclotron
source term, $Q^{\rm cyc}$, using the monochromatic expression
\begin{equation}
Q^{\rm cyc}(z,\epsilon)
\equiv 1.92 \times 10^{52} \, r_0^2 \, \rho^2
\, B_{12}^{-7/2} \, H\left(\epsilon_c \over kT_e\right) \,
e^{-\epsilon_c/kT_e} \, \delta(\epsilon-\epsilon_c)
\ ,
\label{eq7.4}
\end{equation}
obtained by combining equations~(\ref{eq7.1}) and (\ref{eq7.2}).

The cyclotron source term described by equation~(\ref{eq7.4}) is
localized in energy and distributed in space. Following Lyubarskii \&
Sunyaev (1982), we assume that the electrons in the optically thin
plasma above the thermal mound are isothermal, and we also assume that
the magnetic field $B$ has a cylindrical geometry with a constant value
throughout the column. The neglect of the dipole variation of the field
geometry and strength is expected to have a small effect on the emergent
spectrum because most of the radiation is created, reprocessed, and
emitted through the column walls within one or two scattering optical
depths above the stellar surface. Compared with the cylindrical geometry
employed in our model, the inclusion of the actual dipole geometry of
the magnetic field will tend to increase the plasma density at the
bottom of the accretion column, due to the funnel-like shape of the
column walls. This is likely to further increase the concentration of
the radiative processes near the bottom of the column. However, a small
fraction of the photons may nonetheless diffuse to a significant
altitude (perhaps several kilometers) above the star before escaping,
because the scattering cross section $\sigpar$ for photons propagating
parallel to the magnetic field is generally much less than Thomson (see
\S~6.2). This issue requires further exploration using a numerical model
that includes the exact dipole dependence of the magnetic field.

For a constant value of $B$, the spatial variation of the cyclotron
source term reduces to $Q^{\rm cyc} \propto \rho^2$, which reflects the
two-body nature of the collisional excitation process. We can obtain the
particular solution for the column-integrated spectrum of the escaping
radiation resulting from the reprocessing of the cyclotron photons by
using equation~(\ref{eq7.4}) to substitute for $Q$ in
equation~(\ref{eq5.14}). The energy integration is trivial, and we find
after substituting for $B_{12}$ using equation~(\ref{eq2.1}) and
integrating over $z_0$ that the particular solution for the
column-integrated escaping photon spectrum is given in cgs units by
\begin{eqnarray}
\Phi^{\rm cyc}_\epsilon(\epsilon)
&=& {3.43 \times 10^{-16} \dot M \, H(\chi_c) \, \xi^2
\sqrt{\alpha^3 w} \ \epsilon^{\kappa-2}\over
\sigbar \, \epsilon_c^{\kappa+3/2}
e^{(\epsilon_c+\epsilon)/(2kT_e)}}
\ \sum_{n=0}^\infty \ {\Gamma(\mu-\kappa+1/2) \, n! \over
\Gamma(1+2\mu) \, \Gamma(n+1/2)}
\nonumber
\\
\quad
&\times& X_n \, A_n \,
M_{\kappa,\mu}[\min(\chi,\chi_c)] \,
W_{\kappa,\mu}[\max(\chi,\chi_c)]
\ ,
\label{eq7.5}
\end{eqnarray}
where $X_n$ is computed using equation~(\ref{eq5.12}) and we have made the
definitions $\chi \equiv \epsilon/kT_e$ and $\chi_c \equiv \epsilon_c / kT_e$.
The quantity $A_n$ represents the spatial integration
\begin{equation}
A_n \equiv \int_{\taumound}^{\taumax}
e^{3 \alpha \tau_0^2/2} \, g_n(\tau_0) \, \tau_0^{-1} \, d\tau_0
\ ,
\label{eq7.6}
\end{equation}
where $\taumax$ and $\taumound$ denote the optical depths at the top of
the accretion column and the surface of the thermal mound, respectively
(see eqs.~[\ref{eq6.1.2}] and [\ref{eq6.3.5}]). By substituting for
$g_n$ using equation~(\ref{eq4.11}), we find that $A_n$ can be evaluated
analytically, yielding
\begin{equation}
A_n = \sum_{m=0}^n {\Gamma(n+1/2) \, (2w)^m \, (3-w)^{-m} \over
\Gamma(m+1/2) \, 2 \, m! \, (n-m)!}
\left\{\Gamma\left[m,{\alpha (w-3) \taumound^2 \over 4}\right]
- \Gamma\left[m,{\alpha (w-3) \taumax^2 \over 4}\right]
\right\}
\ ,
\label{eq7.7}
\end{equation}
where $\Gamma(a,z)$ denotes the incomplete gamma function.

\subsection{Blackbody Radiation}

Photons are produced at the surface of the thermal mound with a
blackbody distribution, and therefore the surface flux is equal to $\pi$
times the intensity (Rybicki \& Lightman 1979). Following Becker \&
Wolff (2005b), we define the function $S(\epsilon)$ so that
$\epsilon^2 S(\epsilon) \, d\epsilon$ represents the number of photons
emitted from the surface per second in the energy range between
$\epsilon$ and $\epsilon + d\epsilon$. We can relate $S(\epsilon)$ to
the Planck distribution by expressing the amount of energy emitted per
second from the surface of the thermal mound (with area $\pi r_0^2$) in
the energy range between $\epsilon$ and $\epsilon + d\epsilon$ as
\begin{equation}
\epsilon^3 S(\epsilon) \, d\epsilon
= \pi \, r_0^2 \cdot \pi \, B_\epsilon(\epsilon) \, d\epsilon \ ,
\label{eq7.8}
\end{equation}
where
\begin{equation}
B_\epsilon(\epsilon) = {2 \, \epsilon^3 \over c^2 h^3} \, {1
\over e^{\epsilon/k\Tmound} - 1}
\label{eq7.9}
\end{equation}
denotes the blackbody intensity and $\Tmound$ is the temperature of the
gas in the thermal mound. Note that the units for $B_\epsilon$ are ${\rm
ergs \ s^{-1} \, ster^{-1} \, cm^{-2} \, erg^{-1}}$.

The function $S$ is related to the source term $Q$ appearing in the
transport equation~(\ref{eq3.1}) via
\begin{equation}
Q^{\rm bb}(z,\epsilon) \equiv S(\epsilon) \, \delta(z-\zmound)
\ ,
\label{eq7.10}
\end{equation}
which can be combined with equations~(\ref{eq7.8}) and (\ref{eq7.9})
to conclude that
\begin{equation}
Q^{\rm bb}(z,\epsilon) = {2 \, \pi^2 r_0^2 \over c^2 h^3}
\, {\delta(z-\zmound) \over e^{\epsilon/k\Tmound} - 1}
\ .
\label{eq7.11}
\end{equation}
This result for the blackbody source term is localized in physical
space, but it possesses a distributed (continuum) energy dependence,
which is the exact opposite of the cyclotron source given by
equation~(\ref{eq7.4}).

We can obtain the particular solution for the column-integrated spectrum
of the escaping radiation resulting from the reprocessing of the
blackbody photons by using equation~(\ref{eq7.11}) to substitute for
$Q$ in equation~(\ref{eq5.14}). In this case the spatial integration is
trivial, and we find after some algebra that
\begin{equation}
\Phi^{\rm bb}_\epsilon(\epsilon) = {2 \, \pi^2 r_0^2 \over c^2 h^3}
\int_0^\infty
{\greencolumn(\zmound,\epsilon_0,\epsilon) \over \dot N_0} \
{\epsilon_0^2 \, d\epsilon_0 \over e^{\epsilon_0/k\Tmound}-1}
\ ,
\label{eq7.12}
\end{equation}
where $\greencolumn$ is computed using equation~(\ref{eq5.10}).

\subsection{Bremsstrahlung Radiation}

The bremsstrahlung emission spectrum produced by electrons streaming
along a strong magnetic field is modified from the standard thermal
result due to the appearance of the resonance at the cyclotron energy
$\epsilon_c$ (Lauer et al. 1983). However, it can be shown that the
angle-averaged emission spectrum obtained by integrating over the
one-dimensional Maxwellian electron distribution in the accretion column
agrees fairly well with the field-free thermal expression for photon
energies $\epsilon < \epsilon_c$. We can therefore describe the
bremsstrahlung photon source using the standard thermal formula in
combination with a separate (resonant) cyclotron term described by
equation~(\ref{eq7.2}). Based on equation~(5.14b) from Rybicki \&
Lightman (1979), we find that the thermal component of the
bremsstrahlung (free-free) photon production rate per unit volume per
unit energy is given in cgs units by
\begin{equation}
\dot n_\epsilon^{\rm ff} = 3.68 \times 10^{36} \, \rho^2 \,
T_e^{-1/2} \, \epsilon^{-1} \, e^{-\epsilon/kT_e}
\ ,
\label{eq7.13}
\end{equation}
for a plasma composed of pure, fully-ionized hydrogen. Equation~(\ref{eq7.13})
can be combined with equation~(\ref{eq7.1}) to define
the free-free source term
\begin{equation}
Q^{\rm ff}(z,\epsilon)
\equiv 1.16 \times 10^{37} \, r_0^2 \, \rho^2
\, T_e^{-1/2} \, \epsilon^{-3} \, e^{-\epsilon/kT_e}
\ , \ \ \ \ \ 
\zmound < z < \zmax
\ ,
\label{eq7.14}
\end{equation}
where $\zmax$ and $\zmound$ denote the altitudes at the top of the
accretion column and the surface of the thermal mound, respectively (see
\S~6.1). Equation~(\ref{eq7.14}) describes the spectrum of the injected,
optically-thin bremsstrahlung emission for photons with energy
$\epsilonabs < \epsilon < \epsilon_c$, where $\epsilonabs$ represents
the thermal self-absorption cutoff, below which the spectrum becomes
optically thick. The calculation of $\epsilonabs$ is complicated by the
fact that the soft photons created via bremsstrahlung emission may be
upscattered due to collisions with fast electrons before they are
reabsorbed. The energy amplification associated with electron scattering
tends to enhance the survival probability of the photons because the
absorption length is a rapidly increasing function of the photon energy.
The effective value of $\epsilonabs$ can therefore be estimated by
setting $\ell_{\rm sc} = 1/\alpha^{\rm ff}_\nu$, where $\ell_{\rm
sc}=(\sig \, n_e)^{-1}$ is the electron scattering mean free path and
$\alpha^{\rm ff}_\nu$ is the free-free absorption coefficient given by
equation~(5.18b) from Rybicki \& Lightman (1979). This yields
\begin{equation}
{\epsilonabs \over k T_e} = 6.08 \times 10^{12} \, T_e^{-7/4} \, \rho^{1/2}
\ .
\label{eq7.14b}
\end{equation}
The utilization of the Thomson cross section is justified because most
of the bremsstrahlung photons are emitted roughly perpendicular to the
magnetic field due to the dipole nature of the radiation pattern (e.g.,
Lauer et al. 1983). In our applications, the value adopted for
$\epsilonabs$ is obtained by averaging equation~(\ref{eq7.14b}) over the
vertical structure of the column. We typically find that $0.01 \lapprox
\epsilonabs/kT_e \lapprox 0.1$.

The bremsstrahlung source term given by equation~(\ref{eq7.14}) possesses
distributed spatial and energy dependences, and therefore it is more
challenging to implement than the cyclotron and blackbody sources given
by equations~(\ref{eq7.4}) and (\ref{eq7.11}), respectively. By using
equation~(\ref{eq7.14}) to substitute for $Q$ in equation~(\ref{eq5.14}),
we find after several steps that the particular solution for the
column-integrated spectrum of the escaping radiation resulting from the
reprocessing of the bremsstrahlung photons can be written in cgs units as
\begin{eqnarray}
\Phi^{\rm ff}_\epsilon(\epsilon)
&=& {2.80 \times 10^{-12} \, \dot M \, \xi^2 \sqrt{\alpha^3 w}
\ \epsilon^{\kappa-2} \, e^{-\epsilon/(2kT_e)} \over
\sigbar \, (kT_e)^{\kappa+1/2}}
\nonumber
\\
&\phantom{a}&
\quad
\times \ \sum_{n=0}^\infty \ {\Gamma(\mu-\kappa+1/2) \, n! \over
\Gamma(1+2\mu) \, \Gamma(n+1/2)} \ X_n \, A_n \, B_n
\ ,
\label{eq7.15}
\end{eqnarray}
where $X_n$ and $A_n$ are computed using equations~(\ref{eq5.12})
and (\ref{eq7.7}), respectively. The quantity $B_n$ represents the
energy integral
\begin{equation}
B_n \equiv \int_{\chiabs}^\infty \chi_0^{-1-\kappa}
\, e^{-\chi_0/2} \, M_{\kappa,\mu}[\min(\chi,\chi_0)] \,
W_{\kappa,\mu}[\max(\chi,\chi_0)]
\, d\chi_0
\ ,
\label{eq7.16}
\end{equation}
where $\chi\equiv \epsilon/kT_e$, $\chi_0\equiv \epsilon_0 / kT_e$,
and $\chiabs \equiv \epsilonabs/kT_e$. This integral can be evaluated
analytically, but the result is very complex and therefore we do not state
it here.

\section{ASTROPHYSICAL APPLICATIONS}

By combining results from the previous sections in the paper, we can now
compute the spectrum emitted from an X-ray pulsar accretion column due
to the bulk and thermal Comptonization of bremsstrahlung, cyclotron, and
blackbody seed photons. The theoretical prediction for the
phase-averaged photon count rate spectrum measured at Earth is computed
using
\begin{equation}
F_\epsilon(\epsilon) \equiv {\columntotal(\epsilon) \over 4 \pi D^2}
\ ,
\label{eq8.1}
\end{equation}
where
\begin{equation}
\columntotal(\epsilon) \equiv \left[
\Phi^{\rm cyc}_\epsilon(\epsilon) +
\Phi^{\rm bb}_\epsilon(\epsilon) +
\Phi^{\rm ff}_\epsilon(\epsilon) \right] A_c(\epsilon)
\label{eq8.2}
\end{equation}
denotes the total spectrum of the Comptonized radiation escaping from
the column, $D$ is the distance to the source, and the function $A_c$
represents a Gaussian cyclotron absorption feature given by the form
(e.g., Heindl \& Chakrabarty 1999; Orlandini et al. 1998; Soong et al.
1990)
\begin{equation}
A_c(\epsilon) \equiv 1 - {d_c \over \sigma_c
\sqrt{2 \pi}}
\, e^{-(\epsilon-\epsilon_c)^2
/(2\sigma^2_c)}
\ .
\label{eq8.3}
\end{equation}
The quantities on the right-hand side in equation~(\ref{eq8.2}) express
the contributions to the escaping spectrum due to the reprocessing of
cyclotron radiation ($\Phi^{\rm cyc}_\epsilon$; eq.~[\ref{eq7.5}]),
blackbody radiation ($\Phi^{\rm bb}_\epsilon$; eq.~[\ref{eq7.12}]), and
bremsstrahlung emission ($\Phi^{\rm ff}_\epsilon$; eq.~[\ref{eq7.15}]),
respectively. In this section we will use equation~(\ref{eq8.1}) to
compute observed spectra based on parameters corresponding to three
specific X-ray pulsars. The theoretical spectra will be compared with
the observational data for each source in order to gain insight into the
effects of bulk and thermal Comptonization in the accretion shock on the
formation of the X-ray continuum. The spectral results presented here
are example calculations, rather than detailed quantitative fits, which
will be reported in subsequent papers.

The sources treated here are Her~X-1, LMC~X-4, and Cen~X-3, which
represent several of the brightest known X-ray pulsars. In each case the
values adopted for the stellar mass and radius are $M_* = 1.4 \, \Msun$
and $R_* = 10\,$km, respectively. The accretion dynamics in these
high-luminosity sources is expected to be dominated by radiation
pressure, and therefore we expect that $\xi \approx 2/\sqrt{3}$ (see
Becker 1998). The scattering cross section for photons propagating
perpendicular to the magnetic field in our calculations is given by
$\sigperp=\sig$ (see eq.~[\ref{eq6.2.1}]), the accretion rate $\dot M$
is constrained by the observed X-ray luminosity, and the source distance
$D$ is determined using published estimates. The remaining fundamental
input parameters that must be set in order to calculate the theoretical
spectrum are $\delta$, $B$, $r_0$, and $T_e$, which are listed in
Table~1 for each of the models considered here. Once these quantities
are specified, the additional parameters $\alpha$, $\taumax$, $\sigpar$,
$\sigbar$, $\taumound$, $\Tmound$, $\tautrap$, and $\epsilonabs$ can be
computed using equations~(\ref{eq3.19}), (\ref{eq6.1.2}),
(\ref{eq6.2.2}), (\ref{eq6.2.3}), (\ref{eq6.3.5}), (\ref{eq6.3.8}),
(\ref{eq6.5.6}), and (\ref{eq7.14b}), respectively, with the results
reported in Table~2. Note that in each case, we find that $\taumax
\lapprox \tautrap$, which indicates that the observed emission is
produced in the ``trapped'' region of the accretion column, as expected
(see \S~6.5). Table~2 also includes results for the inflow speed at the
thermal mound surface, $\vmound$ (eq.~[\ref{eq6.3.4}]), and the
accretion mass flux, $J$ (eq.~[\ref{eq6.3.7b}]). In general, we
find in our applications that $\vmound \sim 0.03\,c$, which represents
an enormous deceleration from the free-fall velocity $\vff \sim 0.6\,c$
at the top of the column. Since there is not much variation in the value
of $\vmound$, we expect that the mound temperature $\Tmound$ will
increase with increasing $J$ (see eq.~[\ref{eq6.3.7}]).

Once all of the parameters are specified, the corresponding count-rate
spectrum is evaluated using equation~(\ref{eq8.1}). The spectral results
presented here were computed using the first five terms in the
expansions (eqs.~[\ref{eq7.5}], [\ref{eq7.12}], and [\ref{eq7.15}]),
which generally provide at least three decimal digits of accuracy. As a
check on the numerical results for the spectral components, we confirm
that the number of photons escaping from the column per unit time is
exactly equal to the number injected, as required in the steady-state
scenario considered here. In addition to the Comptonized spectrum
evaluated using equation~(\ref{eq8.1}), the results for the total
spectra displayed in Figures~6, 7, and 8 also include an Fe K$\alpha$
emission line modeled using the Gaussian function
\begin{equation}
F^{\rm K}_\epsilon(\epsilon) \equiv {d_{\rm K} \over \sigma_{\rm K}
\sqrt{2 \pi}} \, e^{-(\epsilon-
\epsilon_{\rm K})^2/(2\sigma^2_{\rm K})}
\ ,
\label{eq8.5}
\end{equation}
where $d_{\rm K}$ is the associated total photon flux measured at Earth.
Note that in the approach taken here, this spectral feature is treated
in an ad hoc manner by simply adding it to the final spectrum rather
than by subjecting the iron line photons to subsequent Comptonization.
Unfortunately, a more sophisticated approach is beyond the scope of the
present paper due to the complexity of the spectral formation process
for the iron emission line photons. The associated auxiliary parameters
$\epsilon_{\rm K}$, $d_{\rm K}$, and $\sigma_{\rm K}$ are listed in
Table~3, along with $\epsilon_c$, $d_c$, and $\sigma_c$ (see
eq.~[\ref{eq8.3}]).

\subsection{Her X-1}

In Figure~6 we display the theoretical count-rate spectrum computed
using equation~(\ref{eq8.1}) along with the deconvolved (incident),
phase-averaged {\it BeppoSAX} data reported by Dal Fiume et al. (1998)
for Her~X-1, which has the instrumental response removed. The values
used for the fundamental theory parameters correspond to model~1, with
$\xi=1.45$, $\sigperp=\sig$, $\delta=1.80$, $B=3.80 \times 10^{12}\,$G,
$r_0=44\,$m, $T_e=6.25 \times 10^7\,$K, $\dot M=1.11 \times 10^{17}\,\rm
g\,s^{-1}$, and $D=5\,$kpc, as indicated in Table~1. The associated
values for the computed parameters $\alpha$, $\taumax$, $\sigpar$,
$\sigbar$, $\taumound$, and $\Tmound$ are reported in Table~2, and the
additional auxiliary parameters are listed in Table~3. Results are
presented for the total spectrum, as well as for the individual
contributions to the observed flux due to the Comptonization of
cyclotron, blackbody, and bremsstrahlung seed photons. The lack of a
strong Wien peak in the spectrum, along with the moderate value of
$\delta$, indicate that thermal Comptonization is unsaturated in this
source. It is interesting to note that the value we obtain for $r_0$ is
about an order of magnitude below the upper limit for the column radius
computed using equation~(\ref{eq6.6.3}), which yields $r_0 \lapprox
462\,$m.

Although the results presented here are not fits to the data, it is
interesting to note that the general shape of the pulsar spectrum
predicted by the theory agrees very well with the observations for Her
X-1, including both the quasi-exponential cutoff at high energies and
the power law shape at lower energies. In the case of Her X-1, we find
that reprocessed (Comptonized) blackbody emission from the thermal mound
makes a relatively small contribution to the spectrum due to the small
radius of the accretion column. We also find that reprocessed cyclotron
emission is swamped by reprocessed bremsstrahlung due to the high
temperature of the infalling plasma. Hence we conclude that the X-ray
spectrum emitted by Her X-1 is completely dominated by Comptonized
bremsstrahlung emission.

\begin{figure}[t]
\begin{center}
\epsfig{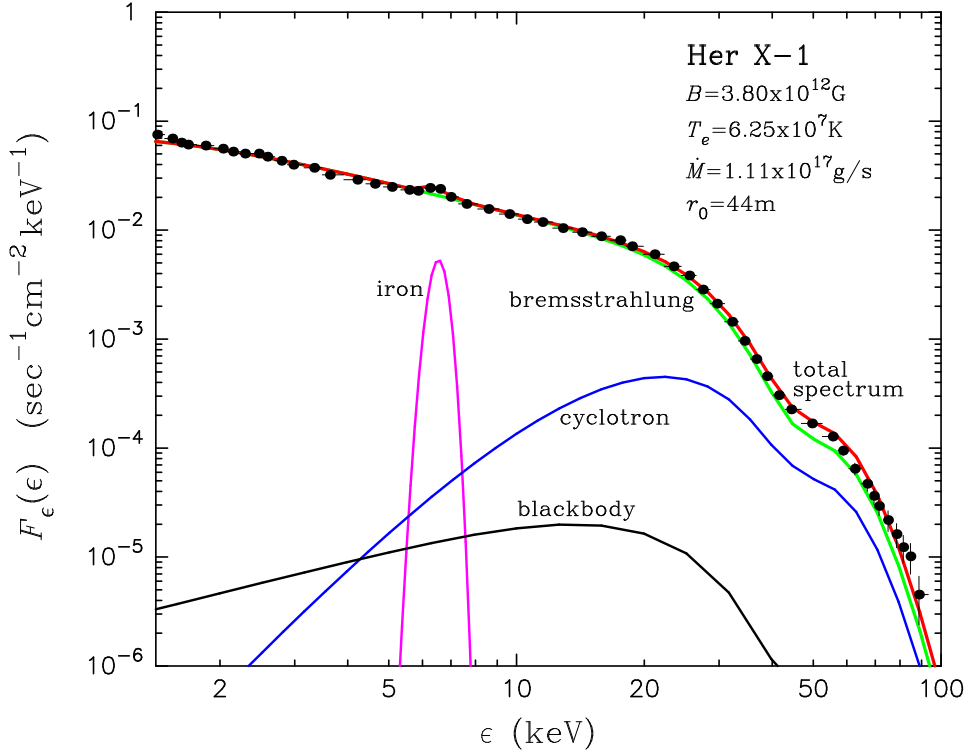}
\end{center}
\caption{Theoretical column-integrated count rate spectrum
$F_\epsilon(\epsilon)$ evaluated using eq.~(\ref{eq8.1}) based on the
model~1 parameters listed in Table~1, compared with the deconvolved
(incident) X-ray spectrum for Her X-1 reported by Dal Fiume et al.
(1998; {\it circles and crosses}). The plots include the total spectrum
as well as the individual components due to Comptonized bremsstrahlung
radiation, Comptonized cyclotron emission, Comptonized blackbody
radiation, and the iron emission line feature, as indicated. Note that
the total spectrum is strongly dominated by the bremsstrahlung
component.}
\end{figure}

\subsection{LMC X-4}

In Figure~7 we compare the theoretical count-rate spectrum computed
using equation~(\ref{eq8.1}) with the deconvolved, phase-averaged {\it
BeppoSAX} data reported by La Barbera et al. (2001) for LMC~X-4. The
values used for the fundamental theory parameters correspond to model~2,
with $\xi=1.15$, $\sigperp=\sig$, $\delta=1.30$, $B=3.28 \times
10^{12}\,$G, $r_0=680\,$m, $T_e=5.90 \times 10^7\,$K, $\dot M=2.00
\times 10^{18}\,\rm g\,s^{-1}$, and $D=55\,$kpc (see Table~1). The
associated values for the computed and auxiliary parameters are listed
in Tables~2 and 3, respectively. Note that the results for the accretion
rate $\dot M$ and the column radius $r_0$ are much larger than those for
Her X-1, although the electron temperatures are nearly the same. In this
case, reprocessed cyclotron emission makes a slightly stronger
contribution to the high-energy spectrum than in model~1, although the
X-ray spectrum is clearly dominated by Comptonized bremsstrahlung
emission. In model~2, the effect of thermal Comptonization (relative to
bulk) is somewhat stronger than in model~1, as indicated by the smaller
value of $\deltapar$ (see eq.~[\ref{eq6.4.4}]). This effect tends to
flatten the LMC X-4 spectrum in the 2-22~keV energy range as compared
with the Her X-1 spectrum, although the thermal Comptonization is
clearly unsaturated, as demonstrated by the absence of a Wien peak in
the spectrum. We also note that the reprocessed blackbody emission makes
a much smaller contribution to the observed spectrum in model~2 than in
model~1 because the thermal mound temperature $\Tmound$ has dropped
substantially, due to the marked decrease in the mass flux, $J$, as
indicated in Table~2. In contrast to the Her X-1 case, for LMC X-4 we
find that the value obtained for $r_0$ is comparable to the upper limit
for the column radius computed using equation~(\ref{eq6.6.3}), which
gives $r_0 \lapprox 727\,$m. This suggests a possible connection
between the column radius and the luminosity in accretion-powered X-ray
pulsars that can be further explored using parameter studies based on a
large number of sources.

\begin{figure}[t]
\begin{center}
\epsfig{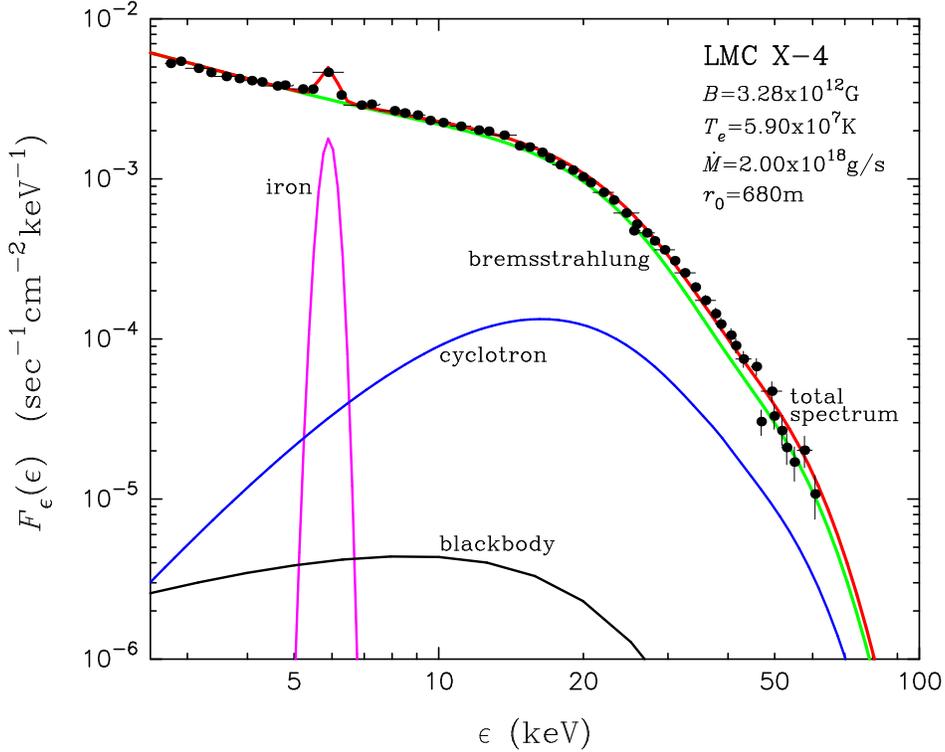}
\end{center}
\caption{Same as Fig.~6, except the data correspond to LMC X-4
and the theoretical spectra were computed based on the model~2
parameters reported in Table~1. The data were reported by La Barbera
et al. (2001; {\it circles and crosses}).}
\end{figure}

\subsection{Cen X-3}

In Figure~8 we plot the deconvolved, phase-averaged {\it BeppoSAX} data
reported by Burderi et al. (2000) for Cen~X-3 along with the theoretical
count-rate spectrum computed using equation~(\ref{eq8.1}). The values
adopted here for the fundamental theory parameters correspond to
model~3, with $\xi=1.25$, $\sigperp=\sig$, $\delta=3.71$, $B=2.63 \times
10^{12}\,$G, $r_0=730\,$m, $T_e=3.40 \times 10^7\,$K, $\dot M=1.51
\times 10^{18}\,\rm g\,s^{-1}$, and $D=8\,$kpc (see Table~1), and the
additional parameters are listed in Tables~2 and 3. In this case only,
interstellar absorption has been included based on a hydrogen column
density of $N_{\rm H}=2.0 \times 10^{22}\,{\rm cm}^{-2}$ in order to
reproduce the observed low-energy turnover (cf. Burderi et al. 2000). We
note that the values obtained for the accretion rate $\dot M$ and the
column radius $r_0$ are once again much larger than those for Her X-1.
It is apparent that the contributions due to Comptonized cyclotron and
blackbody emission are completely negligible, and the observed spectrum
is dominated by Comptonized bremsstrahlung photons.

The effect of bulk Comptonization is more pronounced in this case than
in the spectra of Her X-1, as indicated by the larger value obtained for
$\deltapar$. We also note that the values obtained for the accretion
flux $J$ and the thermal mound temperature $\Tmound$ in model~3 are
lower than those associated with models~1 and 2. However, despite the
decrease in the mound temperature, the Comptonized blackbody emission
has increased in strength to become roughly comparable to the
reprocessed cyclotron emission, which reflects the decrease in the
magnetic field strength $B$. As in the case of LMC X-4, we find that the
value for $r_0$ associated with Cen X-3 is comparable to the upper limit
for the column radius computed using equation~(\ref{eq6.6.3}), which
yields $r_0 \lapprox 744\,$m. This supports the possibility of a
connection between the column radius and the X-ray luminosity, as
suggested in \S~8.2.

\section{CONCLUSIONS}

\begin{figure}[t]
\begin{center}
\epsfig{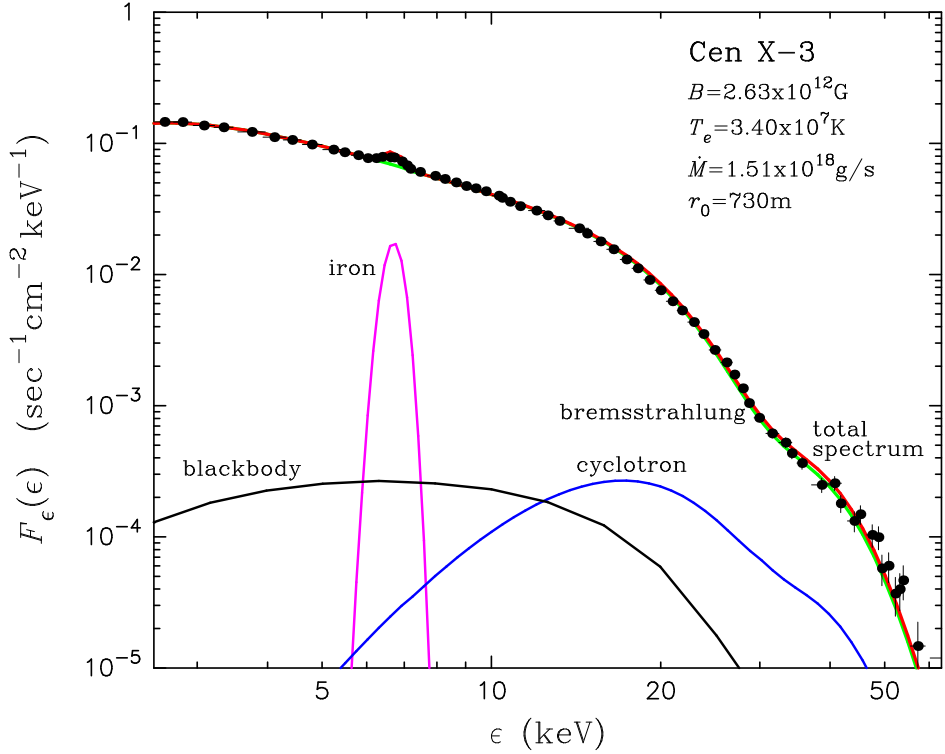}
\end{center}
\caption{Same as Fig.~6, except the data correspond to Cen X-3 and the
theoretical spectra were computed based on the model~3 parameters
reported in Table~1. The data were reported by Burderi et al. (2000;
{\it circles and crosses}). This case also includes interstellar absorption
with hydrogen column density $N_{\rm H}=2.0 \times 10^{22}\,{\rm cm}^{-2}$.}
\end{figure}

The new theoretical model developed here represents the first complete
calculation of the X-ray spectrum associated with the physical
accretion scenario first suggested by Davidson (1973), including bulk
and thermal Comptonization. For given values of the stellar mass $M_*$
and the stellar radius $R_*$, our model has six fundamental free
parameters, namely the accretion rate $\dot M$, column radius $r_0$, the
electron temperature $T_e$, the magnetic field strength $B$, the photon
diffusion parameter $\xi$ (eq.~[\ref{eq3.12}]), and the Comptonization
parameter $\deltapar$ (eq.~[\ref{eq4.5}]). The accretion rate is
constrained by the observed luminosity, which effectively removes this
parameter from the set. A unique solution for the remaining free
parameters can be found by comparing the model spectrum with the
observations for a particular source. The analytical nature of the
new model makes it amenable to incorporation into standard analysis
packages such as XSPEC.

We have shown that the combination of bulk and thermal Comptonization
naturally leads to the development of emergent spectra in
accretion-powered X-ray pulsars that are in good agreement with the
observational data for three bright sources. The main result of the
derivation is the analytical solution for the column-integrated Green's
function describing the escaping radiation spectrum, given by
equation~(\ref{eq5.10}). Based on this closed-form solution, we have
developed expressions for the emergent spectra resulting from the
reprocessing of seed photons injected as a result of cyclotron,
blackbody, and bremsstrahlung emission, given by
equations~(\ref{eq7.5}), (\ref{eq7.12}), and (\ref{eq7.15}),
respectively (we note that additional seed photons may also be supplied
via external illumination of the column, which is not considered here).
In each case, we confirm that the number of photons escaping from the
column per unit time is exactly equal to the number injected, as
expected in the steady-state scenario considered here. Our main results,
summarized in Figures~6, 7, and 8, indicate that the spectra observed
from the bright sources Her X-1, LMC X-4, and Cen X-3 are dominated by
Comptonized bremsstrahlung.

The treatment of the full energy and angle dependence of the scattering
cross sections for the two polarization modes computed using
equations~(\ref{eq2.2}) and (\ref{eq2.4}) is not possible using an
analytical approach. As an alternative, in this paper we have adopted a
two-dimensional approximation by modeling the scattering of photons
propagating either parallel or perpendicular to the magnetic field using
the energy- and mode-averaged expressions given by
equations~(\ref{eq2.6}) and (\ref{eq2.7}), respectively. The results we
have obtained for the directional components of the cross section listed
in Table~2 are not unreasonable, given the uncertainties in the detailed
physical properties of the accretion column. In particular, we find that
the cross section for photons propagating parallel to the field
direction, $\sigpar$, is substantially reduced relative to the Thomson
value. This result is consistent with equation~(\ref{eq2.6}), since
$\epsilon \ll \epsilon_c$ for most of the photons in typical X-ray
pulsar spectra. Furthermore, the cross section values reported in
Table~2 confirm that $\sigpar \ll \sigbar \ll \sigperp$, as expected
(see \S~6.2 and Canuto et al. 1971). Compared to the behavior obtained
in our model, we expect that utilization of the exact cross sections
given by equations~(\ref{eq2.2}) and (\ref{eq2.4}) would tend to
increase the efficiency of thermal Comptonization, relative to the bulk
process, for photons with energy $\epsilon$ close to the cyclotron
energy $\epsilon_c$. This effect is therefore expected to cause some
modification of the parameter values required to fit the observed X-ray
pulsar spectra. Since the radiative transfer problem including the exact
cross sections is not tractable analytically, further study of this
issue awaits the development of a complete numerical model, which we
intend to pursue in future work.

The validity of the approximate velocity profile employed here
(eq.~[\ref{eq3.15}]) was established in Figure~5 by demonstrating
excellent agreement between the bulk Comptonization spectrum computed
using our new model in the limit $T_e \to 0$ and the corresponding
spectrum obtained using the pure bulk Comptonization model of Becker \&
Wolff (2005b), which is based on the exact velocity profile
(eq.~[\ref{eq3.16}]). The values obtained for the accretion column
radius $r_0$ (reported in Table~1) comply with the corresponding upper
limits computed using equation~(\ref{eq6.6.3}), which was derived based
on the detailed dynamics of neutron star accretion flows (Lamb et al.
1973; Harding et al. 1984). We find that $r_0$ is essentially equal to
the upper limit for the two brightest sources considered here (Cen X-3
and LMC X-4), while it is about an order of magnitude smaller than this
limit for Her X-1. Since the luminosity of Her X-1 is roughly an order
of magnitude lower then the other two sources, this suggests a possible
connection between the luminosity and the column size that should be
further investigated by developing numerical models that simultaneously
treat the dynamics and the spectral formation in X-ray pulsars.

The transfer of energy from the gas to the photons is generally
dominated by bulk Comptonization in accretion-powered X-ray pulsars.
However, our results confirm that unsaturated thermal Comptonization
nonetheless plays a significant role in the spectral formation process.
Specifically, we find that the thermal process effectively transfers
energy from high to low frequency radiation, which contributes to the
observed quasi-exponential cutoffs at high energies and at the same time
causes flattening of the spectrum at lower energies. While it has been
known for some time that pulsars generally have spectra that are well
fitted using a combination of a power-law plus a quasi-exponential
cutoff, until now this form has been adopted in a purely ad hoc manner.
The new model described here finally provides a firm theoretical
foundation for this empirical result, based on model parameters that are
directly tied to the physical properties of the source. It therefore
represents a significant step in the development of a comprehensive
theory for the spectral formation process in accretion-powered X-ray
pulsars.

The predictions of our analytical model can be tested both
observationally and also by developing numerical models that allow the
relaxation of several of the key assumptions made here. For example,
within the context of a numerical model one can include the exact
spatial variation of the shape and strength of the dipole magnetic
field, as well as the full energy and angle dependence of the electron
scattering cross section. Although these effects are beyond the scope of
the present paper, we nonetheless believe that our main conclusions
regarding the spectral shape will be verified by more detailed models.

The authors would like to thank Ken Wolfram, Paul Ray, Lev Titarchuk,
and Kent Wood for a number of stimulating conversations. The authors are
also grateful to the anonymous referee who provided a number of
insightful comments that led to significant improvements in the
manuscript. PAB would also like to acknowledge generous support provided
by the Office of Naval Research.


\clearpage

\begin{deluxetable}{clcccccrr}
\tabletypesize{\scriptsize}
\tablecaption{Input Model Parameters\label{tbl-1}}
\tablewidth{0pt}
\tablehead{
\colhead{Model}
& \colhead{Object}
& \colhead{$\xi$}
& \colhead{$\delta$}
& \colhead{$B$ (G)}
& \colhead{$\dot M$ (g~s$^{-1}$)}
& \colhead{$T_e$ (K)}
& \colhead{$r_0$ (m)}
& \colhead{$D$ (kpc)}
}
\startdata
1
&Her X-1
&1.45
&1.80
&$3.80 \times 10^{12}$
&$1.11 \times 10^{17}$
&$6.25 \times 10^7$
&44.0
&5.0
\\
2
&LMC X-4
&1.15
&1.30
&$3.28 \times 10^{12}$
&$2.00 \times 10^{18}$
&$5.90 \times 10^7$
&680.0
&55.0
\\
3
&Cen X-3
&1.25
&3.71
&$2.63 \times 10^{12}$
&$1.51 \times 10^{18}$
&$3.40 \times 10^7$
&730.0
&8.0
\\
\enddata


\end{deluxetable}



\begin{deluxetable}{ccccccccccc}
\tabletypesize{\scriptsize}
\tablecaption{Computed Model Parameters\label{tbl-2}}
\tablewidth{0pt}
\tablehead{
\colhead{Model}
& \colhead{$\alpha$}
& \colhead{$\sigpar/\sig$}
& \colhead{$\sigbar/\sig$}
& \colhead{$J$ (g\,s$^{-1}$\,cm$^{-2}$)}
& \colhead{$\Tmound$ (K)}
& \colhead{$\vmound / c$}
& \colhead{$\epsilonabs/kT_e$}
& \colhead{$\taumound$}
& \colhead{$\taumax$}
& \colhead{$\tautrap$}
}
\startdata
1
&0.401
&$4.15 \times 10^{-5}$
&$2.93 \times 10^{-4}$
&$1.83 \times 10^9$
&$5.68 \times 10^7$
&0.027
&0.098
&0.067
&1.37
&1.58
\\
2
&0.319
&$4.85 \times 10^{-5}$
&$3.98 \times 10^{-4}$
&$1.38 \times 10^8$
&$2.91 \times 10^7$
&0.026
&0.034
&0.080
&1.12
&1.77
\\
3
&0.346
&$8.30 \times 10^{-5}$
&$4.51 \times 10^{-4}$
&$9.02 \times 10^7$
&$2.48 \times 10^7$
&0.023
&0.072
&0.066
&1.07
&1.70
\\
\enddata


\end{deluxetable}



\begin{deluxetable}{ccccccc}
\tabletypesize{\scriptsize}
\tablecaption{Auxiliary Parameters\label{tbl-3}}
\tablewidth{0pt}
\tablehead{
\colhead{Model}
& \colhead{$\epsilon_c$ (keV)}
& \colhead{$\sigma_c$ (keV)}
& \colhead{$d_c$ (keV)}
& \colhead{$\epsilon_{\rm K}$ (keV)}
& \colhead{$\sigma_{\rm K}$ (keV)}
& \colhead{$d_{\rm K}$ (s$^{-1}$\,cm$^{-2}$)}
}
\startdata
1
&44.0
&13.0
&25.20
&6.55
&0.30
&$4.00 \times 10^{-3}$
\\
2
&38.0
&21.2
&38.29
&5.90
&0.22
&$1.00 \times 10^{-3}$
\\
3
&30.5
&11.0
&21.50
&6.70
&0.27
&$1.20 \times 10^{-2}$
\\
\enddata


\end{deluxetable}

\end{document}